\documentclass[useAMS,usenatbib]{mn2e}
\usepackage{times}
\usepackage{graphicx}

\newcommand{\kms}   {km~s$^{-1}$}

\newcommand{\eg}    {e.\,g.}
\newcommand{\ie}    {i.\,e.}

\newcommand{\sii}   {[S\,{\sc ii}]}

\newcommand{\fe}    {[Fe\,{\sc ii}]}
\newcommand{\vlsr}  {$V_\rmn{LSR}$}
\newcommand{\vtot}  {$V_\rmn{tot}$}
\newcommand{\vtan}  {$V_\rmn{T}$}

\newcommand{\hmol} {$\mathrm{H}_2~\upsilon=1$--$0~\mathrm{S(1)}$}
\newcommand{\hmo} {$\upsilon=1$--$0~\mathrm{S(0)}$}
\newcommand{\hmu} {$\upsilon=2$--1$~\mathrm{S(1)}$}

\newcommand{\hmj} {$\upsilon=3$--$1~\mathrm{S(3)}$}
\newcommand{\hmjj} {$\upsilon=2$--$0~\mathrm{Q(1)}$}
\newcommand{\hmh} {$\upsilon=1$--$0~\mathrm{S(6)}$}
\newcommand{\hmhh} {$\upsilon=1$--$0~\mathrm{S(9)}$}
\newcommand{\hmolh} {$\upsilon=1$--$0~\mathrm{S(7)}$}

\title[LIRIS MOS of HH~223]
{3-D Kinematics of the near-IR HH 223 outflow in L723}

\author[R. L\'opez et al.]
{R. L\'opez,$^{1}$\thanks{E-mail:
rosario,robert@am.ub.es;
jap,bgacia@iac.es;
gabriel.gomez@gtc.iac.es
}
J. A. Acosta-Pulido,$^{2,3}$\footnotemark[1]
R. Estalella,$^{1}$\footnotemark[1]
G. G\'omez$^{4,2}$\footnotemark[1], and
\newauthor
B. Garc\'\i a-Lorenzo$^{2,3}$\footnotemark[1]
\\
$^{1}$Departament d'Astronomia i Meteorologia (IEEC-UB), Institut de Ci\`encias
del Cosmos, U. de Barcelona, Mart\'{\i} i Franqu\`es 1, 
E-08028 Barcelona, Spain\\
$^{2}$Instituto de Astrof\'{\i}sica de Canarias, E-38200 La Laguna, Spain\\
$^{3}$Departamento de Astrof\'{\i}sica, Universidad de La Laguna, E-38205,
Tenerife, Spain\\
$^{4}$GTC Project Office, GRANTECAN S.A. (CALP), E-38712 Bre\~na Baja, La Palma,
Spain.\\
}

\begin{document}

\date{Accepted 2014 December 4. Received 2014 December 1; in original form 2014
November 12}

\pagerange{\pageref{firstpage}--\pageref{lastpage}} \pubyear{2014}

\maketitle

\label{firstpage}

\begin{abstract} 

In this  work we derive the full 3-D kinematics of the near-infrared outflow HH~223,
located in the dark cloud Lynds 723 (L723), where   a well-defined quadrupolar CO outflow
is found.  HH~223  appears projected onto the two lobes of the east-west CO outflow. The
radio continuum source VLA~2, towards the centre of the CO outflow, harbours  a multiple
system of low-mass young stellar objects. One of the components has been proposed to be
the  exciting source of the east-west CO outflow.

From the analisys of the kinematics, we get further evidence on the relationship between
the near-infrared and CO outflows and on the  location of their exciting source.  The
proper motions were derived  using multi-epoch, 
narrow-band H$_2$ (2.122~$\mu$m line) images. Radial velocities were derived  from the
2.122~$\mu$m line of the spectra. Because of the extended ($\sim$~5~arcmin), S-shaped
morphology of the target, the spectra were obtained with the Multi-Object-Spectroscopy
(MOS) observing mode using the instrument LIRIS at the 4.2~m William  Herschel Telescope.
To our knowledge, this work is the first time that MOS observing mode has been
successfully used in the near infrared range for an  extended target.  

\end{abstract}


\section{INTRODUCTION}

The isolated dark cloud Lynds 723 (L723), located at a distance of 300~$\pm$150~pc
(\citealp{Gol84}) is one of the few sites where a well-defined quadrupolar CO outflow has
been reported (two separate pairs of red-blue lobes; \citealp{Lee02} and references
therein). The 3.6~cm radio continuum source VLA~2 (\citealp{Ang96}), located towards the
centre of the CO outflow, harbours the low-mass source that powers the  outflow.
\citet{Car08} and \citet{Gir09} report that VLA~2 is a multiple system of four (VLA~2A,
2B, 2C and 2D) young stellar objects (YSOs), and that one of the components (VLA~2A) is
powering the east-west CO outflow.  Later on, \citet{Gir09} detect dust emission at
1.35~mm, resolved into two components (SMA~1 and SMA~2). SMA~2, in a more evolved
stage, is harbouring the multiple low-mass protostellar system VLA~2. In addition, they
report emission from the SiO 5--4 line towards the SMA sources, which shows an elongated
morphology that follows the direction of the east-west CO outflow near the exciting source
as well as that of the
near-infrared and optical outflows reported in L723.

\begin{figure*}
\begin{center}
\begin{minipage}{170mm}
\includegraphics[angle=-90,width=160mm]{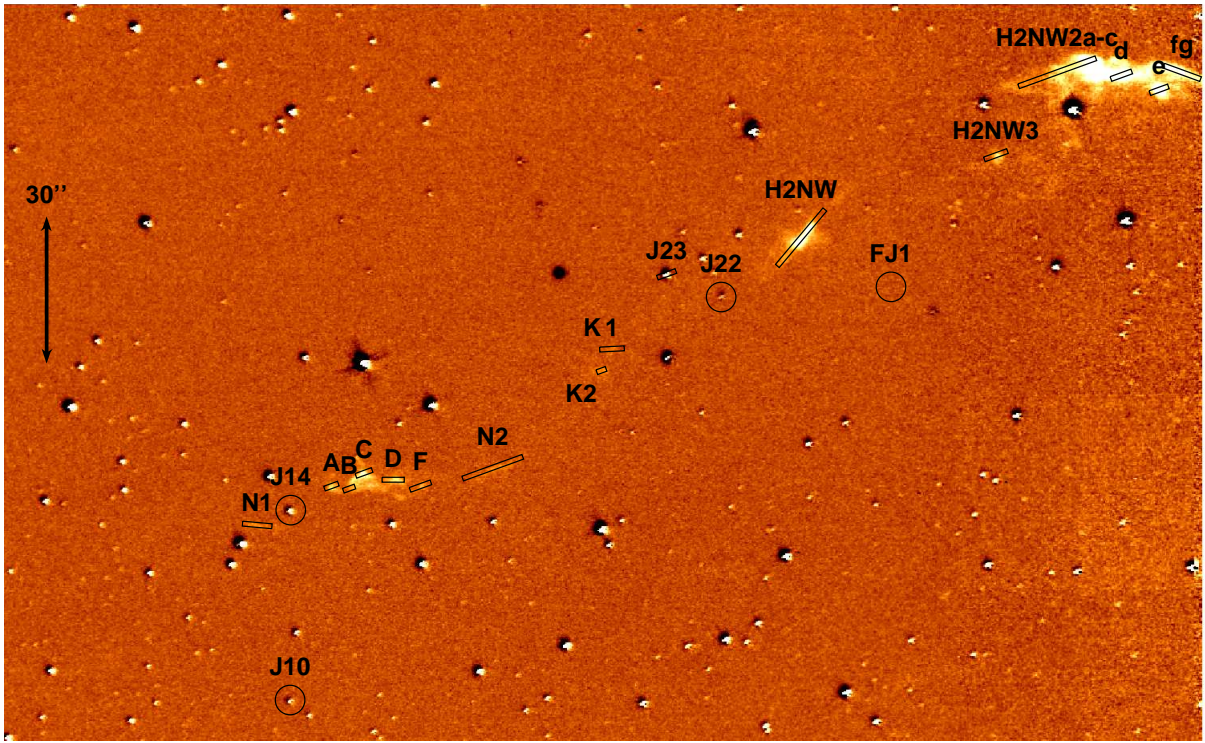}
\caption{Image of the L723 field obtained wiht LIRIS through the H$_2$ 2.122
$\mu$m line filter (the continuum has been substracted). The rectangles
represent the position of the slitlets of the mask designed for 
the  Multi-Object Spectroscopy (MOS) observations. The knots of HH~223 enclosed
in the corresponding slitlet of the mask have been labeled according to the nomenclature of \citet{Lop10}. 
An additional slitlet,
labeled J23 was included to check the individual frame quality. The four
circular appertures, labeled J10, J14, J22 and FJ1, used for mask centering, are also
drawn. North is up and east is to the left.
\label{slit}}
\end{minipage}
\end{center}
\end{figure*}

The protostellar system embedded in SMA~2 seems to be  associated with the 
large-scale outflows detected in other wavelength ranges, as  proposed
from previous deep narrow-band images of the L723 field. In the optical
range, it drives the Herbig-Haro (HH) object 223 of the Reipurth 
Catalogue\footnote{Reipurth, B.: 1999, A general catalogue of Herbig-Haro
objects, 2.\ edition, http://casa.colorado.edu/hhcat}
first detected by \citet{Vrb86} as a ``linear emission
feature'', and later resolved into several knots by \citet{Lop06}.  The
spectra of these knots are characteristic of shock-excited gas, as
revealed from long-slit spectroscopy (\citealp{Lop09}). The emission of
the knots show a complex pattern, both in its kinematics and physical
conditions, as derived from Integral Field Spectroscopy (IFS) observations
by \citet{Lop12}.  In the near-infrared range, \citet{Pal99} detect in a
{\it K}-band image  several H$_2$ emission nebulae, located at both sides
of VLA~2 . Later on, \citet{Lop10} obtain deep images of the L723 field
through narrow-band filters centred on the \fe\ 1.644 $\mu$m and H$_2$
2.122 $\mu$m
lines. The images show a set of H$_2$ emission features,
distributed from the southeast to the northwest of the L723 field,
extending along $\sim$ 5~arcmin ($\sim$ 0.5~pc for a distance of 300~pc).
The H$_2$ emission features are found projected onto the lobes of the\
east-west CO outflow, with a S-shaped morphology, and are proposed
to form part of a H$_2$ outflow,  which also has optical counterpart at
the regions  with low visual extinction, given rise to the HH object 223.
In contrast,  the \fe\ 1.644 $\mu$m image only shows emission associated
with the HH object 223, which could be tracing the densest,
high-ionized and lower extinction  region of the outflow.

The 3-D kinematics
(\ie\ including proper motions and radial velocities) is a more robust tool
than the outflow morphology for a
reliable identification  of the driving source of the outflow: the direction of the
motions of the outflow features points to the position of the driving
source. However,  the kinematics of the near-infrared
HH~223 outflow remained unknown up to date.  With the aim of  obtaining further 
evidence on the relationship between the H$_2$ and the CO outflows of
L723, as well as to check the location of the source driving  the
optical/near-infrared/millimetre outflows, we conducted an in-depth study 
of the kinematics of
the near-infrared outflow. We derived proper motions and radial velocities
of the near-infrared emission features, extending at both sides of the
radio-continuum sources. The proper motions were found from multi-epoch imaging
of the  L723 field, obtained with the same instrumental configuration.
The radial velocity field was derived from {\it K}-band spectroscopy, using the
bright H$_2$ 2.122 $\mu$m line. Because of the extended S-shape morphology
of the H$_2$ emission,  the Multi-Object-Spectroscopy (MOS) observing mode
results more efficient than long-slit mapping if we want to cover most
emission features along the outflow in a reasonable amount of observing time. 
The instrument LIRIS at the WHT offers this 
observing mode in the near-infrared range. 
LIRIS-MOS observing mode
has been widely used to get spectra of  point-sources in crowded fields,
although this is the 
first time  that it has been used to collect spectra of extended  targets like 
the nebular emission
features in the field of L723. The use on extended targets introduces some 
complexities,
such as the slit mask design and the data reduction and calibration procedures
to be followed.

\section{OBSERVATIONS, DATA REDUCTION AND CALIBRATION}

\begin{table}
\caption{Observation Log}
\begin{tabular}{clrr}  
\multicolumn{4}{c}{Imaging}\\ 
\hline
Date  & Grism & Exp. Time & Seeing \\
     &       & (s)       & (arcsec)\\
\hline 
2006-07-20 & H$_2$ & 1800 & 0.8 \\
2010-06-22 & H$_2$ & 1800 &  1\\
2012-07-26$^{a}$ & H$_2$ & 3100 &  1\\
\hline
\multicolumn{4}{c}{$^a$ Obtained under Director's Discretionary Time}\\
\multicolumn{4}{c}{of Spain's Instituto de Astrof\'isica de Canarias}
\end{tabular}
\medskip
\vspace{0.3cm}
\begin{tabular}{clcr}  
\multicolumn{4}{c}{MOS Spectroscopy} \\
\hline
Date  & Grism & Exp. Time$^b$ & Seeing \\
     &       & (s)       & (arcsec)\\
\hline 
2009-08-29 & HR$_K$ & 6$\times$600 & 0.7 \\
2010-06-22 & HR$_J$ & 6$\times$600 & 1 \\
2010-06-22 & HR$_H$ & 6$\times$600 & 1 \\
\hline
\multicolumn{4}{l}{$^b$ On-source integration time}
\end{tabular}
\label{ta:obslog}
\end{table}

Imaging and Multi-Object Spectroscopy observations were made with
the instrument LIRIS (Long-Slit Intermediate Resolution Infrared
Spectrograph; \citealt{Aco03}; \citealt{Man04}) at the 4.2~m Williams Herschel Telescope
(WHT) of the Observatorio del Roque de los Muchachos (ORM, La Palma,
Spain). LIRIS is equipped with a Rockwell Hawaii 1024 $\times$ 1024 HgCdTe array
detector. The spatial scale is 0.25~arcsec~pixel$^{-1}$, giving an image
field of view (FOV) of 4.27 $\times$ 4.27 arcmin$^2$.  
Imaging and spectroscopic data were
processed using the  package {\it
lirisdr}\footnote{http://www.iac.es/project/LIRIS} developed by the LIRIS 
team within the  {\small IRAF}\footnote{{\small IRAF} is distributed by
the National Optical Astronomy Observatories, which are operated by the
Association of Universities for Research in Astronomy, Inc., under
cooperative agreement with the National Science Foundation.} environment.

\begin{table}
\begin{minipage}{\columnwidth}
\caption{Configuration of the mask}
\begin{tabular}{l@{\extracolsep{-1pt}}rrrrr} 
\hline
Knot
&Slitlet 
&Length
&\multicolumn{2}{c}{Center Position}
&PA\\
HH~223-
&ID
&(arcsec)
&$\alpha_{2000}$
&$\delta_{2000}$
&(deg)\\
\hline
N1       & s1	 & 6.0    &19 17 59.034 &+19 11 42.10 &$-95$\\
A        & s2	 & 3.0   &19 17 57.970 &+19 11 50.02 &$-70$\\
B        & s3	 & 2.5  &19 17 57.717 &+19 11 49.51 &$-70$\\
C        & s4	 & 3.5  &19 17 57.501 &+19 11 52.71 &$-70 $\\
D-E      & s5	 & 4.5  &19 17 57.087 &+19 11 51.33 &$-90$\\
F        & s6	 & 4.5  &19 17 56.697 &+19 11 50.04 &$-70$\\
N2       & s7	 & 13.0   &19 17 55.660 &+19 11 53.80 &$-70 $\\
K2       & s8	 & 2.0    &19 17 54.100 &+19 12 13.60 &$-70$\\
K1       & s9	 & 5.0    &19 17 53.950 &+19 12 18.00 &$-87$\\
star J23 & s10	 &4.0  &19 17 53.170 &+19 12 33.20 &$-75$\\
H2-NWab  &s11	 &15.0   &19 17 51.235 &+19 12 40.67 &$-40  $\\
H2-NW3   &s12	 &5.0    &19 17 48.440 &+19 12 57.50 &$-70  $\\
H2-NW2abc&s13	 &16.7 &19 17 47.560 &+19 13 14.40 &$-70   $\\
H2-NW2d  &s14	 &4.5  &19 17 46.643 &+19 13 13.81 &$-70  $\\
H2-NW2e  &s15	 &4.0    &19 17 46.101 &+19 13 10.85 &$-70  $\\
H2-NW2fg &s16	 &8.0    &19 17 45.771 &+19 13 14.40 &$-110  $\\
\hline
\end{tabular}
\label{tslit}
\end{minipage}
\end{table}
 
\subsection{Imaging}

Deep narrow-band images of the L723 field through a filter centred on the H$_2$ 2.122 $\mu$m
line were obtained at three different epochs (2006 July 20, 2010 June 22 and 2012
July 26). The observing strategy consisted of a
5-point dithering pattern. Due to the elongated  morphology of the target, we used
a E-W offset three times larger than the one used  along the N-S direction. The
reduction process included sky subtraction,  flat-fielding, correction of
geometrical distortion, and finally combination of frames using the common
``shift-and-add'' technique.  This final step consisted in dedithering and
co-addition of frames taken at different dither points to obtain a mosaic
covering a field of $\sim~5~\times$~5~arcmin$^2$, which includes  the  HH~223
H$_2$ outflow. 
Astrometric calibration in each  final image was made using the
coordinates from the 2MASS All Sky Catalogue of ten field stars  well
distributed on the observed field. The rms of the transformation was
0.04~arcsec  in both coordinates.

\subsection{Multi Object Spectroscopy}

We obtained spectra of the line-emitting nebulosities forming the HH~223 H$_2$
outflow, which extends over 5.5~arcmin  (equivalent to $\sim$~0.5~pc for a
distance of 300~pc) PA~ following a S~pattern with a 
postion angle (PA)~$\simeq 110 ^\circ$.
The
identification of the H$_2$  filaments and clumps can be found in \citet{Lop10}.
In order to be more efficient for collecting the spectra, we used the  
MOS mode of LIRIS. The observations were performed at two
different  epochs (see Table~\ref{ta:obslog} for details),  using the same MOS
mask at the same nominal positioning. The  FOV of LIRIS 
together with  the particularly elongated  morphology of
the outflow permit to cover most of the  line emitting regions with the slitlets
relatively well  aligned (see Fig.~\ref{slit}). This fact guarantees a rather
consistent wavelength coverage for most of them.  The designed mask  had 16
slitlets: 15 of them were located to cover the outflow features, and one more
(number 10) was positioned onto a star, which was used as a reference target. The
aim of using a reference target was twofold: to register the offset between
different frames, and to assess individual frame
quality.  In addition, the mask design included holes, placed at the position of
three relatively bright  field stars, to support the pointing and
acquisition process of the field.   Each  slitlet was  1 arcsec wide, and its
length ranged from $\sim$ 1.5 to 16 arcsec, depending on the morphology of the
corresponding feature to be sampled  (see  Fig.~\ref{slit} and Table \ref{tslit}).
The orientation of each slitlet was slightly different, which introduces  relative
tilting of
the spectral lines among the spectra of different slitlets. Note that this mask
was the first one designed with the unusual requirement to have slitlets with
different relative orientations. 

\begin{figure}
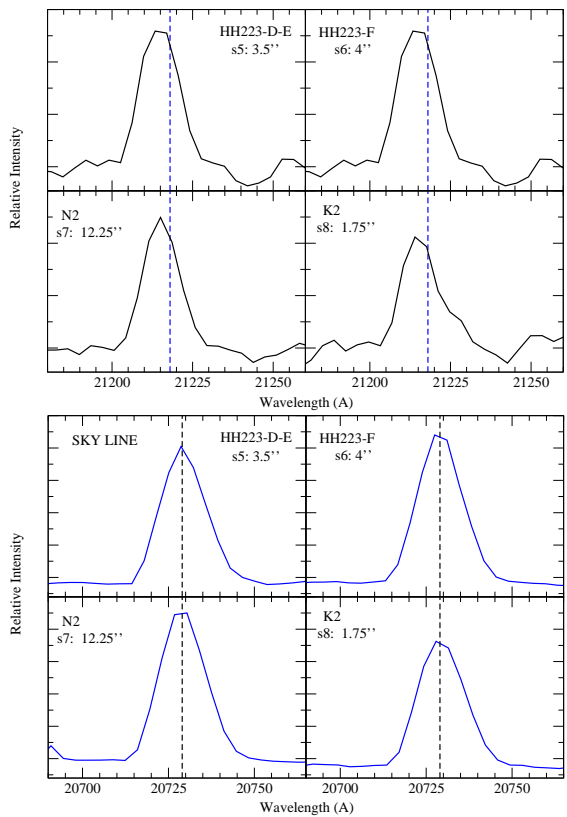

\includegraphics[width=74mm,clip]{fig2a.eps}
\includegraphics[width=74mm,clip]{fig2b.eps}
\caption{Profiles of the  H$_2$ line emission at H$_2$ 2.122 $\mu$m (top) and of the 
 OH sky line at  2.073 $\mu$m (bottom) for the spectra averaged within the slitlets labeled in the
panel. Dashed lines indicate the position that corresponds to the rest wavelength.
\label{exem}} 
\end{figure}

Spectra were obtained using the designed mask and the 3 medium-resolution
(R $\simeq$ 2500) grisms, covering the nominal spectral ranges from 
1.18 to 1.35 $\mu$m, 
from 1.53 to 1.79 $\mu$m, and from 2.07 to 2.44 $\mu$m, in
the {\it J}, {\it H} and  {\it K} spectral windows, respectively.  Total
on-source integration time was 1~hr,  split in individual exposures of
600~s to avoid saturation by the bright sky lines and to compensate for their
time variability. The observations were perfomed repeating the sequence
Object-Sky (OSOS). The sky exposures were obtained in an empty field,
offset by $\sim$ 1 arcmin, roughly  perpendicularly to  the outflow axis, 
since the observed field is crowded and our targets are extended and close
to each other.

The data reduction process of this dataset was rather complex and combined
{\it lirisdr}/{\small IRAF} standard procedures with more dedicated correction steps
developed in {\small IDL}. For the standard procedures we used  the routines
available for the LIRIS MOS mode in the {\it lirisdr} package.   The first step
was to determine and correct the geometrical distortions along the spatial axis
in order  to have the spectra aligned within the CCD rows. Dome white lamp (dome flats,
hereafter) spectra  were used to determine the correction. Next, we used the
a-priori mask design information to trace the slitlet positions and their limits from
a   distortion-free dome--flat frame.  Then the 2-D spectra from each slitlet
were extracted following the limits determined previously and were calibrated 
independentely.  The wavelength calibration was determined from argon and xenon
lamp exposures obtained with the same  instrument configuration through the MOS
mask.  The arc spectra for each slitlet were extracted in the  way described
before. The initial wavelength calibration was determined as in the ``long-slit
case'' following the usual tasks in  {\small IRAF} (i.e. combining {\it identify,
reidentify} and {\it fitcoords}).   The accuracy of the wavelength calibration
was then checked using a set of bright, isolated OH sky lines  present in the
observed wavelength ranges. It was noted that wavelength offsets were present,
which is expected due to spectrograph flexures. The determination of radial
velocities from emission lines is a crucial aspect of the present work,  
therefore we
needed to refine the wavelength calibration by applying an offset determined from the
wavelength difference observed in the mentioned OH sky lines.  After this
correction was applied, the accuracy reached for the wavelength calibration was
better than 0.35~\AA\  in all the observed wavelength range (\ie\ 4.9 \kms\ for
the H$_2$ 2.122~$\mu$m, and 6.3 \kms\ and 8.3~\kms\ for the 1.644 $\mu$m and
1.257~$\mu$m \fe\ 
emission lines).

A check of the
reliability of the wavelength calibration in the wavelength range close to the line used
to derive the kinematics (H$_2$ 2.122 $\mu$m) can bee seen in Fig.~\ref{exem}, which shows
the line profiles of
this line (top panels) and of the OH 2.073 $\mu$m sky line
(bottom panels) of the spectra obtained by averaging the signal within the full
aperture of the slitlet labeled in each of the panels. The position of the
corresponding rest wavelength has been marked with the dashed vertical line.
Note that the emission  appears shifted from the rest wavelength position in
the case of the H$_2$ 2.122 $\mu$m line, while it appears centred at the rest
wavelength for the sky line, as should be expected. We then concluded that the
wavelength calibration was accurate enough to derive  reliable kinematics of
the outflow emission  using the H$_2$ 2.122 $\mu$m  line.

\begin{figure}
\includegraphics[width=84mm]{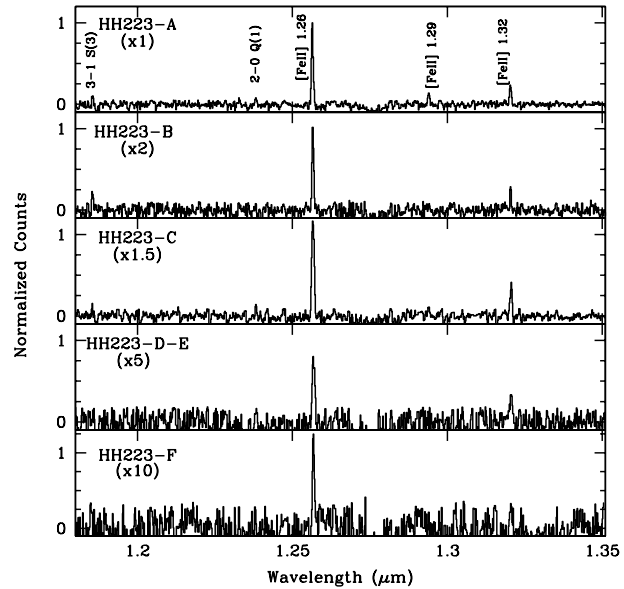}
\caption{{\it J}-band spectra of the slitlets covering the  knots with detected
emission lines of the
 HH~223 H$_2$ outflow, obtained
by averaging the signal within the corresponding slitlet. 
Intensities have been scaled relative to the \fe\ 1.257 $\mu$m
line intensity peak of the HH223-A knot.
\label{intej}}
 \end{figure}

Next we performed the sky subtraction to the science frames using an
average sky spectrum obtained  from the  adjacent sky frames. The resulting 
sky-subtracted spectrum showed large residuals coincident with bright OH  sky
lines.  In order to improve the results we followed an approach similar to that
developed by \citet{Davies07}: first  we separated the  OH emission lines from
the sky continuum. The sky continuum was modelled by a smooth function (a
polynomial  function of degree 4 is  usually a good solution). In order to
scale properly the OH emission spectrum we divided the spectral range  into
segments which correspond to different vibrational transitions of the OH
molecule. We found the scaling  factor for each of the segments that
minimized the difference between the sky  and the object spectra.   Finally,
the sky spectrum was subtracted from the 2-D object extracted spectra, and
we combined them  into an average spectrum for each slitlet. Any possible
offsets along the slit direction were derived from  the relative positions of
the reference stars and then  taken into account before the object frames were
combined.

\section{RESULTS AND DISCUSSION}

\begin{figure}
{\includegraphics[width=83.5mm]{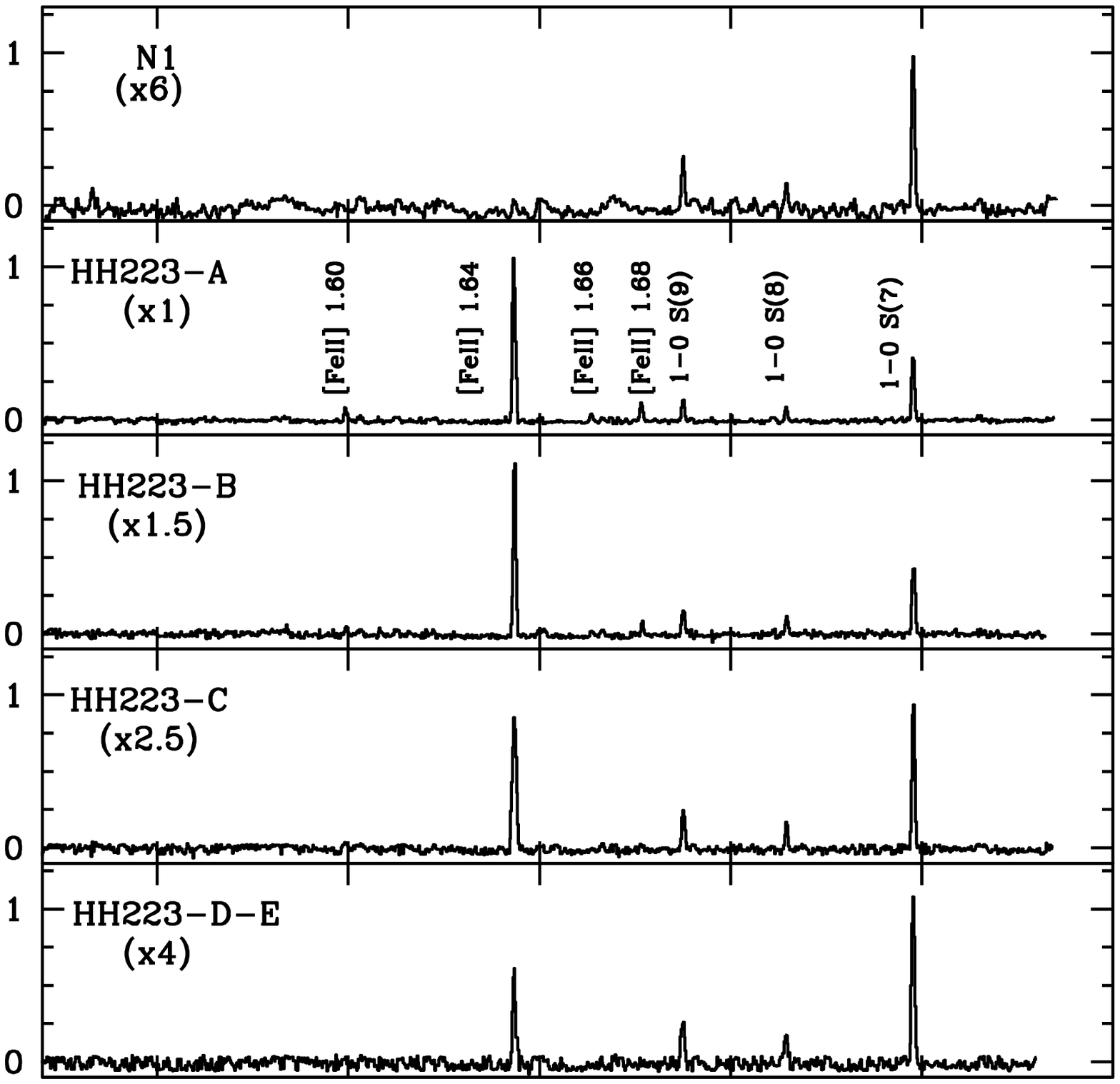}
\vspace{-1.4cm}}
{\vspace{-0.75cm}
\includegraphics[width=83.5mm]{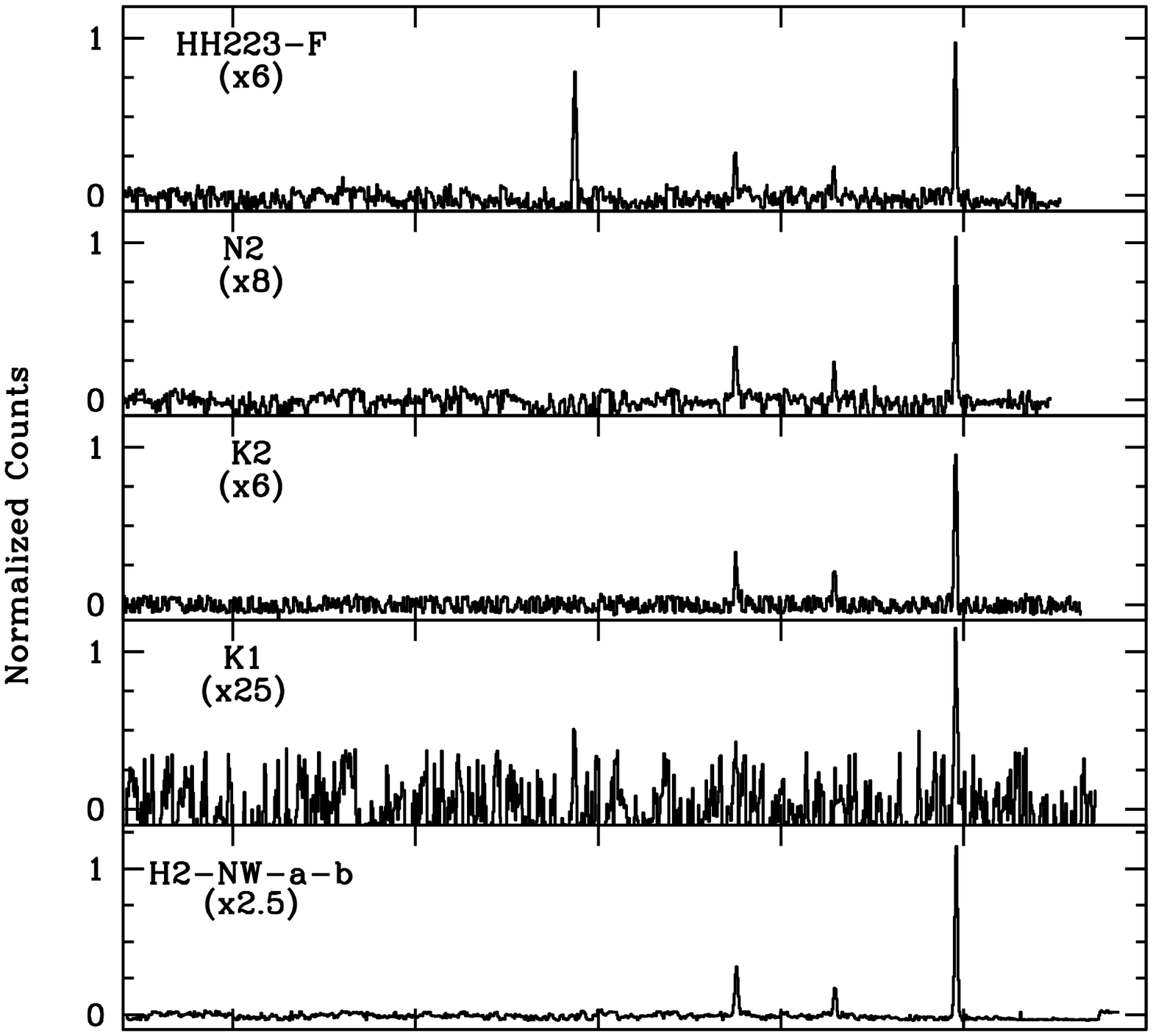}
\vspace{-0.60cm}}
{\vspace{-1.5cm}
\includegraphics[width=83.5mm]{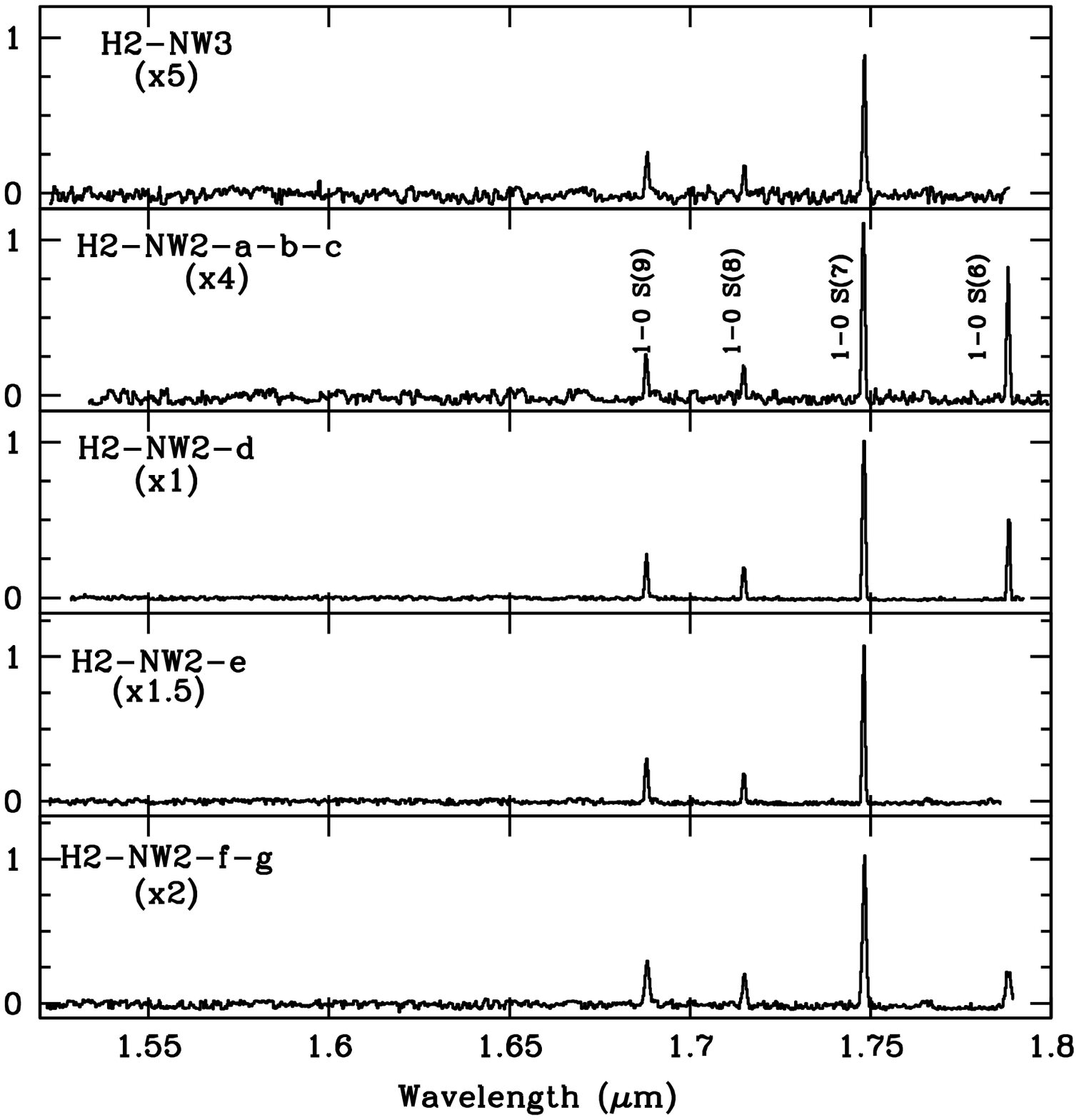}}
\vspace{1cm}
\caption{Same as Fig.~\ref{intej}, but for the {\it H}-band spectra. 
Intensities  have been scaled relative to the H$_2$ 1.75 $\mu$m
line intensity peak of the H2-NW2-d knot.
\label{inteh}}
 \end{figure}

\begin{figure}
{\includegraphics[width=83.5mm]{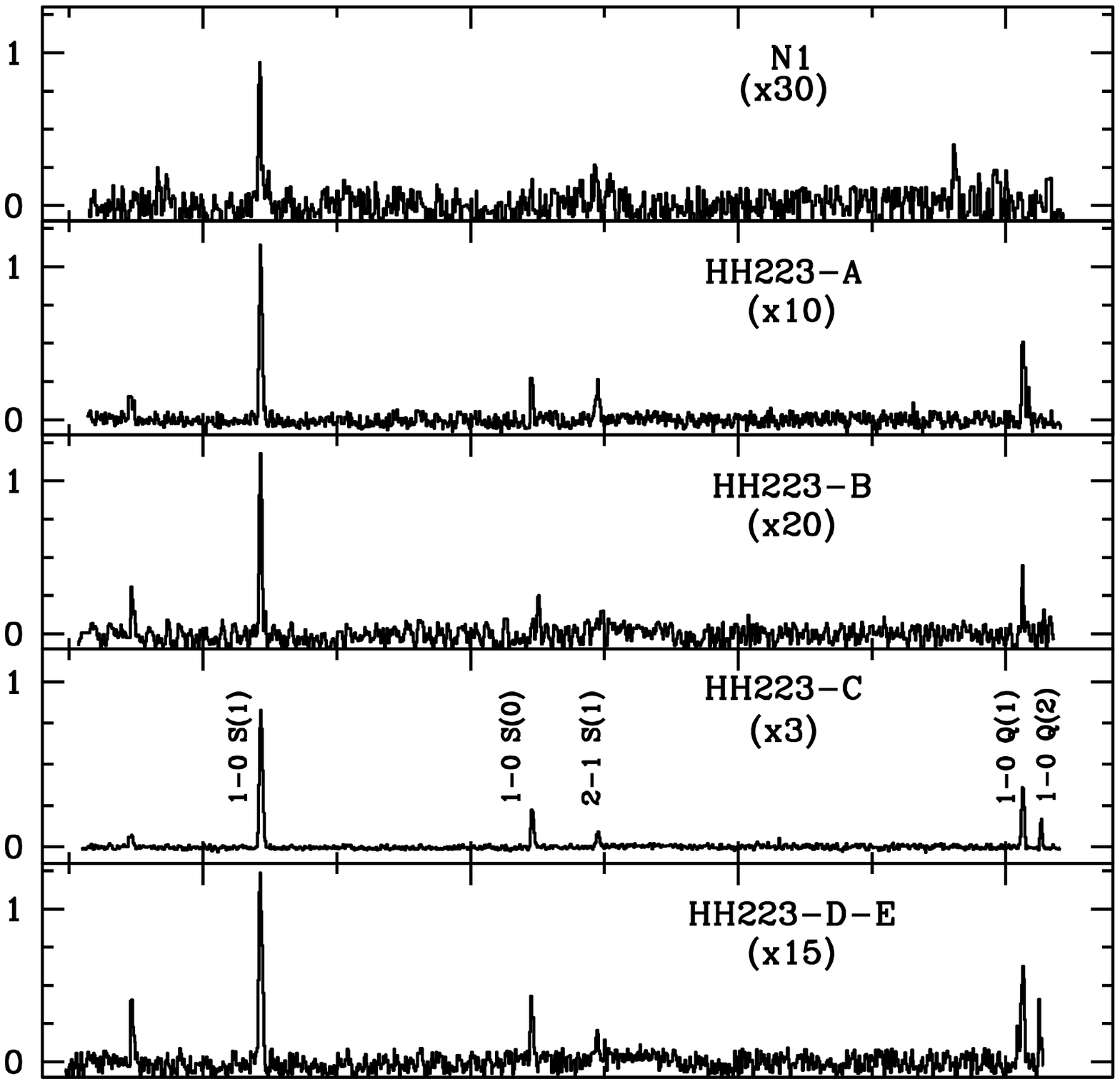}
\vspace{-1.4cm}}
{\vspace{-0.75cm}
\includegraphics[width=83.5mm]{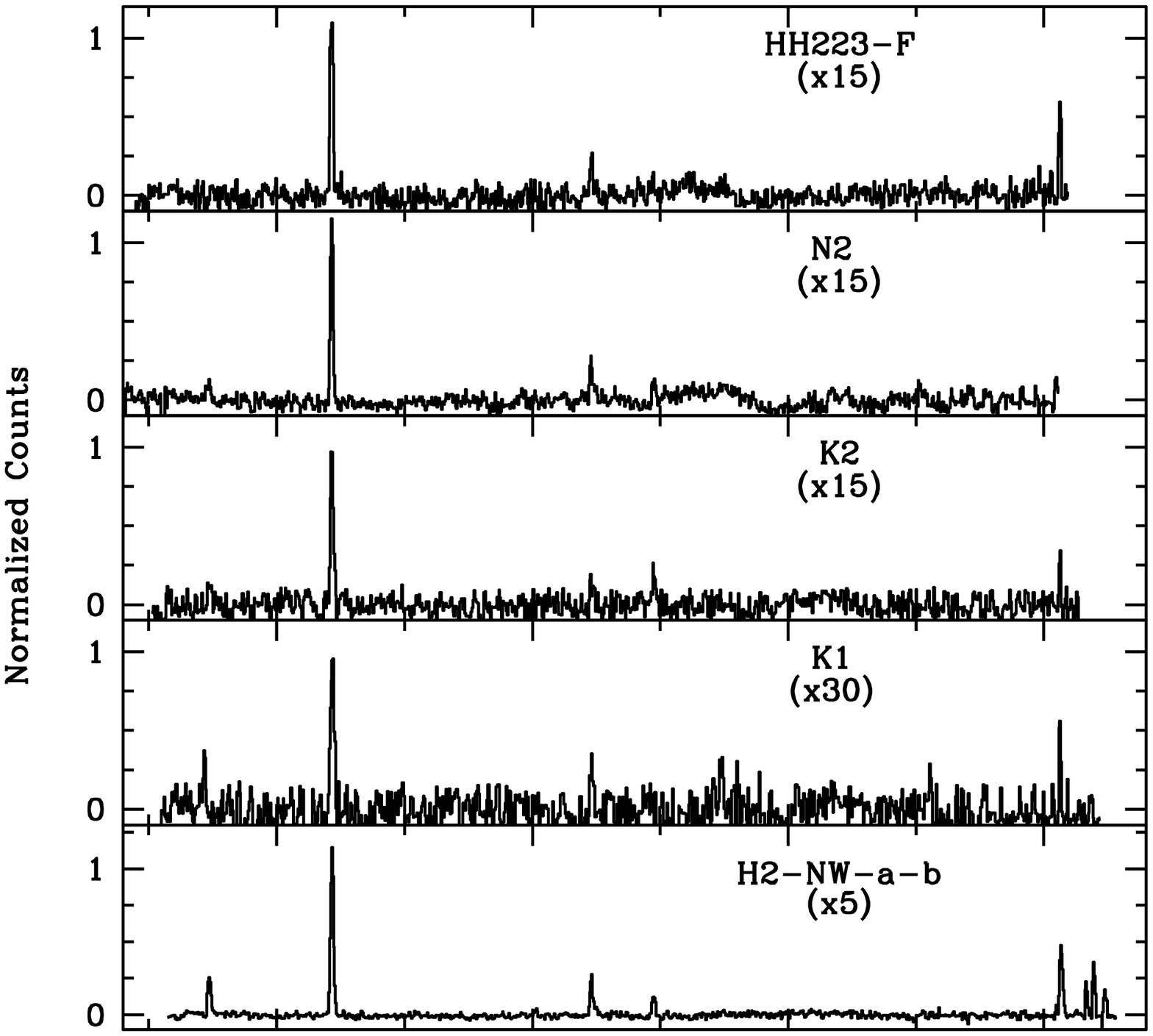}
\vspace{-0.60cm}}
{\vspace{-1.5cm}
\includegraphics[width=83.5mm]{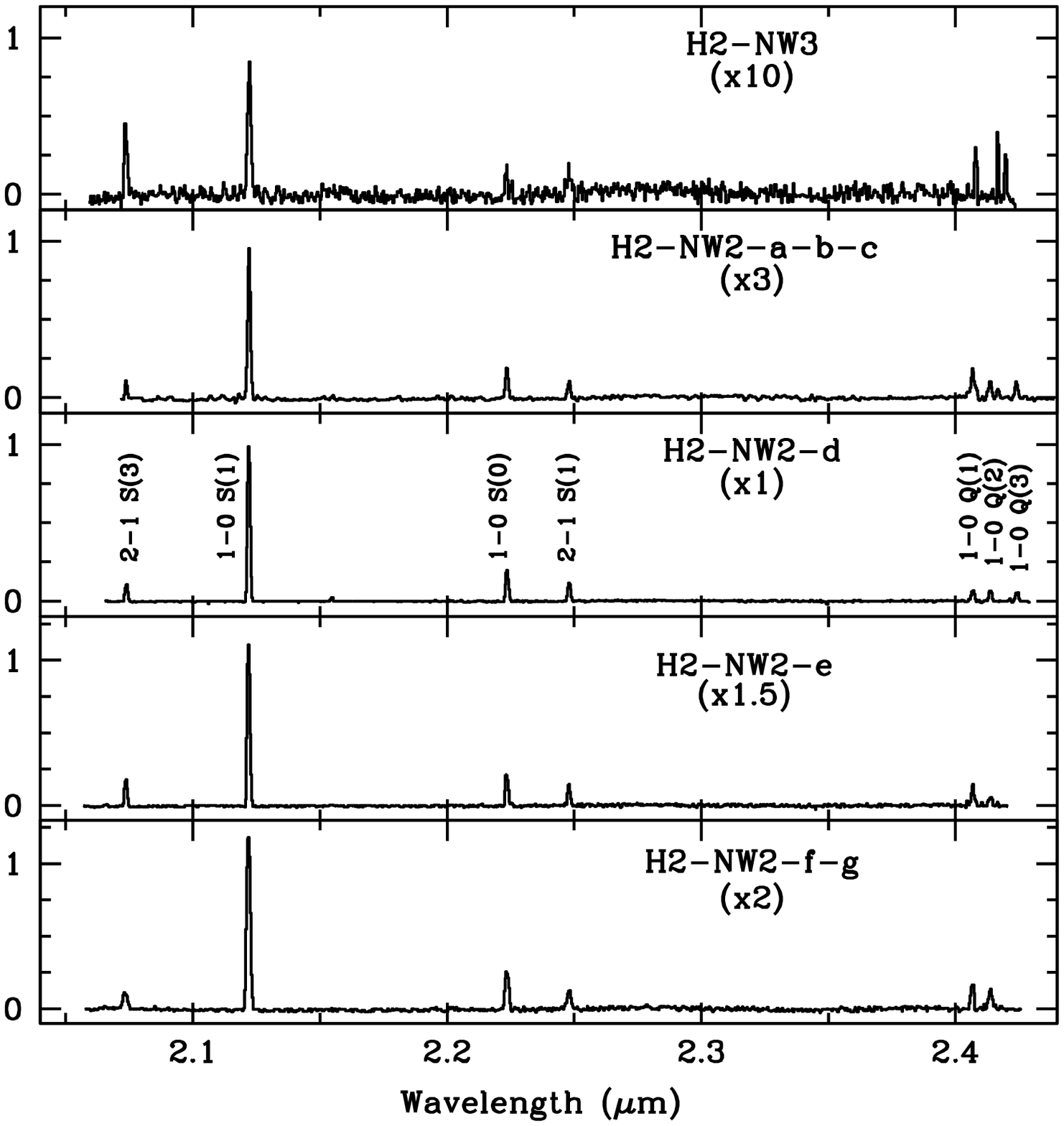}}
\vspace{1cm}
\caption{Same as Fig.~\ref{intej}, but for the {\it K}-band spectra.
Intensities  have been scaled relative to the 2.122 $\mu$m
line intensity peak of the H2-NW2-d knot.
\label{intek}}
 \end{figure}

We obtained the one-dimensional spectrum  for each of the 15 slitlets covering the
nebulosities of the HH~223 H$_2$ outflow  by averaging the signal within the
corresponding slitlet (\ie\  within a rectangular aperture of 1~arcsec  width and
a slit length ranging from $\sim$~2~arcsec to $\sim$~16~arcsec, see Table
\ref{tslit}). The spectra in the {\it J}, {\it H} and {\it K} bands are shown
in Figs.~\ref{intej}, \ref{inteh} and \ref{intek}, respectively. 
In order to search for variations at  scales smaller than
the length of the slitlets, we
extracted, in
addition,  the one-dimensional spectra of each of the H$_2$ features identified
along the HH~223 outflow. The kinematics of HH~223
was derived from the spectra.
The spatial brightness distribution of neutral and ionized gas outflow was also
explored from the 2-D MOS spectra (see Fig~\ref{mapah2}).

\subsection{Emission brightness distribution}

\subsubsection{The neutral outflow}

In the near-infrared, the emission of the neutral  gas of the HH~223 outflow 
is traced by the H$_2$ lines. In the  observed wavelength range,
all the H$_2$ transitions we
detected correspond to those with the lowest excitation levels  ($E \leq$
15000 K).

As can be seen from Figs.~\ref{inteh} and \ref{intek}  the bright H$_2$  \hmolh\ 1.748 $\mu$m
 and \hmol\ 2.122 $\mu$m emission  lines were detected with
high signal-to-noise (SNR $\geq$~10) in all the  spectra. Other H$_2$ transitions
were also detected in several slitlets, and have been marked in the figures. In
the {\it K} band,  the Q-branch emission lines at 2.4 $\mu$m, lying  at the  edge
of the   covered spectral range, were found in most of the slitlets. Additional 
H$_2$ lines, \eg\ the  \hmo\ 2.224 $\mu$m and the \hmu\ 2.248 $\mu$m were found in
several slitlets (these onto the knots  HH~223-A to -F and -H2-NW, -H2-NW2),
but were detected with  a lower  SNR (by a factor $\sim$~5 and $\sim$~10, respectively)
relative to the SNR of the 2.122 $\mu$m). In the {\it H} band, several
H$_2$ transitions  were detected in most of the slitlets  (from the \hmh\ 1.788 $\mu$m
to the \hmhh\ 1.687 $\mu$m lines).  However, only the  bright  1.748 $\mu$m line was
detected at the slitlet S9,  positioned on HH~223-K1, the knot closest to the
site  where the outflow exciting  source is embedded. Finally, in the {\it
J} band,  only the  H$_2$ \hmj\ 1.185 $\mu$m and \hmjj\ 1.238 $\mu$m lines  
were detected (SNR $\simeq$~5) in HH~223-A.  The lack of detection of the {\it
J} band H$_2$ lines in most of the knots  can be due mainly to extinction, 
because  these knots are  more embedded in the cloud. Note that most
of the bright  knots identified in the {\it K} band (\eg\ HH~223-NW and -NW2) lack
of an optical  counterpart. This can be also the case of HH~223-K1. At this position, 
the high extinction could prevent the detection of the other H$_2$ lines lying  at
the  {\it H} band, except the bright 1.748 $\mu$m line.

\begin{figure*}
\includegraphics[width=150mm,clip]{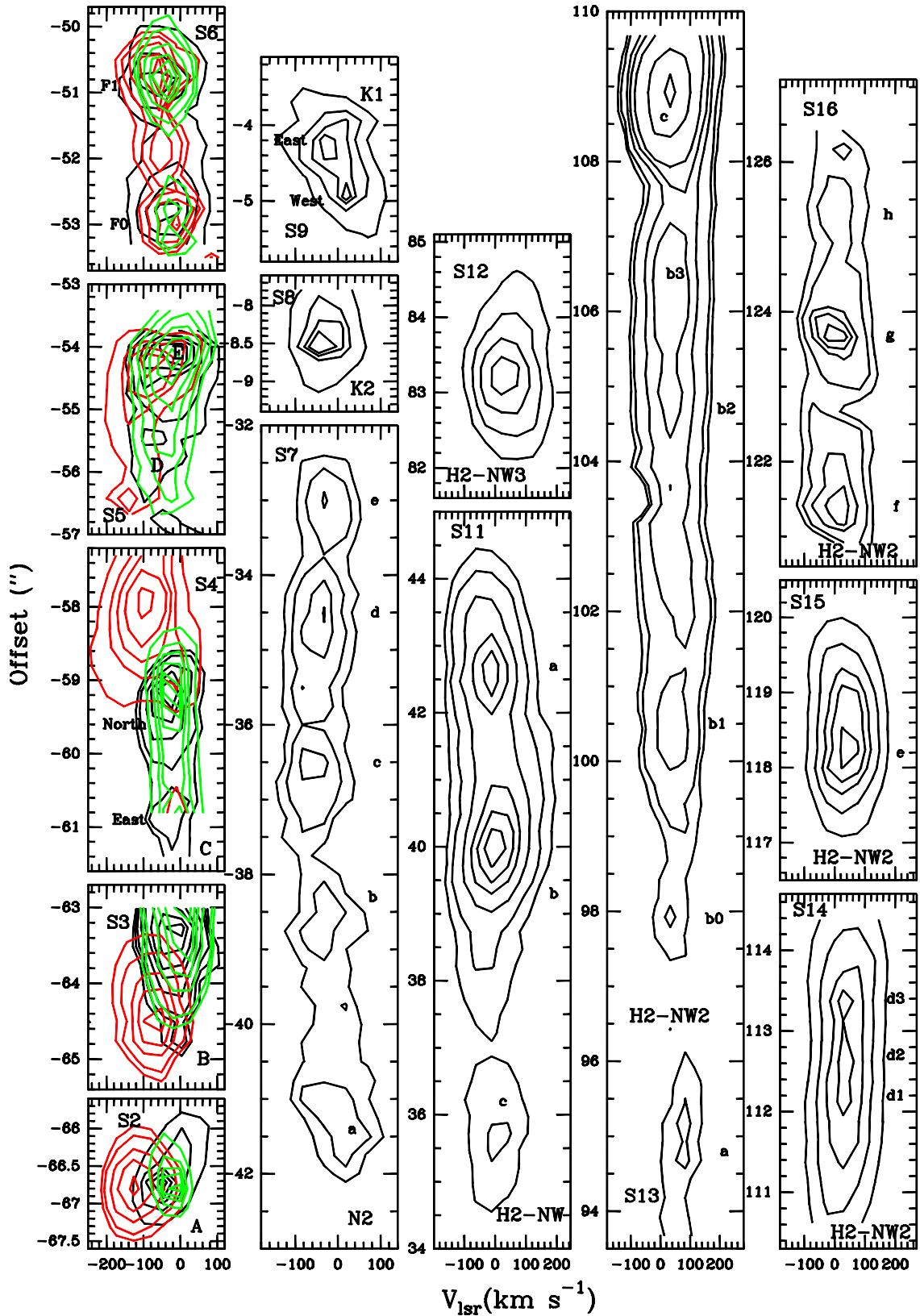}
\caption{Position-velocity maps of the H$_2$ 2.122 $\mu$m line 
 through the slitlet labeled in each panel (black
 contours). The PV maps of the  \fe\ 1.644~$\mu$m are supperposed (red contours) 
 for the slitlets
 where \fe\ emission was detected. The PV maps of the H$_2$ 1.748~$\mu$m line 
 are also displayed (green contours) 
 for these slitlets to check the reliability of the spatial displacement between the 
 H$_2$ and \fe\ emissions.  
 The knots
intersected by each slitlet have been labeled to help with the identification.
\label{mapah2}} 
\end{figure*}

\subsubsection{The ionized outflow}

The emission of the  ionized gas of HH~223 is traced by the \fe\ lines.The \fe\ lines were
detected only through the slitlets positioned on the HH object 223 (i.e., knots
HH~223-A to -F, the knots having an optical counterpart).  The bright
\fe\ 1.257~$\mu$m and 1.644 $\mu$m lines were detected in all these knots. 
Other weaker \fe\ transitions
(\eg\ at 1.600, 1.664 and 1.677 $\mu$m, in the {\it H} band,  and at 1.295 and 1.321
$\mu$m in the  {\it J} band) were only detected in the spectrum of the knot HH~223-A. The
fact that  the  \fe\ transtions  were only detected  in a few knots of the near-infrared
HH~223 outflow is consistent with what was found from previous deep narrow-band {\it
H}-band images of the field (\citealp{Lop10}).

The spatial brightness distribution of
the \fe\ and H$_2$ emissions are not fully coincident, as can be seen from
Fig.~\ref{mapah2} (left panels) by comparing the PV maps of the \fe\ 
1.644~$\mu$m (red contours) and H$_2$ 2.122~$\mu$m (black contours) line emissions.
The greater discrepancies are found for the emissions acquired through the slitlets
S3 and S4, positioned on the knots HH~223-B and HH~223-C. 
In spite of the \fe\
({\it H} band) and H$_2$ 2.122~$\mu$m ({\it K} band) spectra being obtained at two
different observing runs, we want to remark that the observed displacements between
these emissions could not be caused by a difference in the positioning of the
mask on the field in each of the observing runs. To support this assertion, 
Fig.~\ref{mapah2} also displays the
PV map obtained from the {\it H}-band H$_2$ 1.748 $\mu$m line emission
(left panels, green contours). 
In general, the spatial distribution of the emission from 
the 2.122~$\mu$m line appears more extended than the emission from the 
1.748~$\mu$m line, although its spatial brightness distribution 
is closely coincident with that of the  
2.122~$\mu$m line. Hence, the spatial displacement found between the spatial
brightness distribution of the ionized  (\fe) and neutral (H$_2$)
emissions is believed to be reliable.

The knots HH~223-A to -F are far away from the location of the powering source of the
outflow (close to HH~223-K1).  Hence, the origin of the \fe\ emission in these knots cannot
be the gas heated by a close protostar. The most plausible heating mechanism is the
presence of shocks with appropriate strength to ionize and excite the gas outflow. On the
other hand, it is very  unlikely that both the \fe\ and H$_2$ emissions originate from the
same parcel of  shocked gas, because each of these emissions trace shocks with completely
different degree of excitation, which are seen either in projection or are unresolved
within the same beam. It should be noted that the  optical counterpart of the HH~223-A to
HH~223-F knots have revealed to show a rather complex kinematic pattern, as derived from
an IFS mapping with more  complete coverage (\citealp{Lop12}). The IFS results suggest
the presence of a set of shocks of complex morphology,  unresolved with the observed beam
resolution. In this scenario,  the \fe\ emission may be tracing the sites where the
interaction, either between different episodes of mass ejection (like internal working
surfaces) or between the supersonic gas and dense clumps of the wall cavity, is strong
enough to excite the \fe\ transitions. 

We expected to detect \fe\ emission in the spectra obtained
along  the 
HH~223-K1 feature, since the gas would be ionized  given the proximity to the
outflow exciting source. However, we failed to detect \fe\
emission in any of the observed bands. The non-detection of \fe\ emission at this
 position, and for the rest of the HH~223 knots, could be caused by a higher extinction
than in HH~223A-F. The non-detection of optical counterparts for all of these knots favours this
hypothesis.

\subsection{Radial velocities} 

The main aim of this work is to establish the kinematics of the near-infrared HH~223
outflow by deriving  both radial velocity and proper motions along all its knots, covering
$\sim$~5~arcmin in the E-W direction. Because the emission of the  ionized gas (traced by
the \fe\ lines) was only detected in part of the outflow (along $\sim$ 30 arcsec), 
we focused our study on the full
kinematics of the neutral gas, traced by the H$_2$ emission in the {\it K} band, where the
effects of the extinction are lower than in the other two bands, and  the most deeply
embedded knots are detected in the images. In addition, we also derived the radial
velocity of the ionized gas from the brightest \fe\ lines in the {\it J} and {\it
H} bands at the knots  HH~223-A to HH~223-F.

\begin{table}
\caption{\label{vcenslit}
Radial velocities (\vlsr)$^{1}$ of the H$_2$ 2.122 $\mu$m emission within the full slitlet aperture} 
\begin{tabular}{clrr}
\hline
Slitlet &Knot & Offset $^{2}$&  \vlsr $^{3}$\\
        &     & (arcsec)& (\kms)\\
\hline
s1     &N1      &--83.9&  $-72.1\pm 12.4$\\
s2     &A       &--67.0&  $-26.3\pm\phantom{0}9.7$\\
s3     &B       &--64.0&  $-21.3\pm\phantom{0}8.2$\\
s4     &C       &--59.9&  $ -7.0\pm\phantom{0}6.2$\\
s5     &D-E     &--55.4&  $ -33.2\pm 12.4$\\
s6     &F       &--51.4&  $-39.2\pm 14.0$\\
s7     &N2      &--37.7&  $-30.3\pm\phantom{0}6.5$\\
s8     &K2      &--8.3 &  $-24.0\pm\phantom{0}9.8$\\
s9     &K1      &--3.9 &  $  +7.3\pm\phantom{0}8.0$\\
s11    &H2-NWab & 40.3 &  $  +1.5\pm\phantom{0}8.3$\\
s12    &H2-NW3  & 82.8 &  $ +43.3\pm 11.5$\\
s13    &H2-NW2ab&101.9 &  $ +49.0\pm 10.6$\\
s14    &H2-NW2d &112.8 &  $ +36.0\pm\phantom{0}6.1$\\
s15    &H2-NW2e &118.2 &  $ +39.9\pm 12.6$\\
s16    &H2-NW2fg&123.9 &  $ +25.6\pm\phantom{0}7.9$\\ 
\hline 
\end{tabular}
\begin{list}{}
\item
$^{1}$ From Gaussian fits to the H$_2$ 2.122 $\mu$m line of the spectra, averaged for the full 
slitlet apertures (see text).\\
$^{2}$ Offsets from the position of SMA~2 to the centre of the slitlet. 
Typical values of the error are $\leq$ 0.5~arcsec\\
$^{3}$ The error values correspond to the error of the Gaussian fit added in quadrature with
the rms residual of the wavelength calibration.\\
\end{list}
\end{table} 


\begin{figure}
\includegraphics[width=84mm,clip]{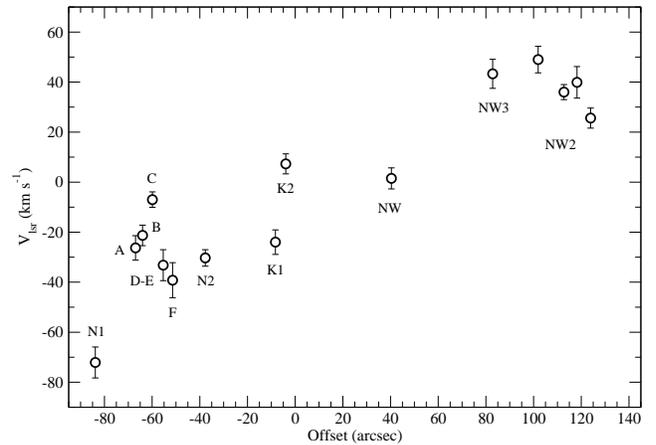}
\caption{Radial velocities
obtained from the line centroid of a  Gaussian fits to the H$_2$ 2.122 $\mu$m
line of the  spectra obtained by averaging the signal within the
full aperture of the corresponding slitlet. Offsets are relative to the position of SMA~2.
\label{censlitav}} 
\end{figure}

In order to derive the  radial velocity field, we obtained
the 
line centroid from a Gaussian fit to the H$_2$ 2.122 $\mu$m 
line.  Note that this line is well separated from any bright OH sky
line. Hence, any residual sky line resulting from an imperfect sky subtraction
does not affect the velocity measurements. Furthermore, 
this line is the brightest one
in the observed {\it K} spectral range, and it is also the only line  
detected in all the knot spectra with a high SNR
(ranging from 10 to 150). As mentioned before, there are other H$_2$  emission 
lines lying in the observed
spectral range, but they do not seem so suitable to trace the outflow
kinematics. The weaker  2.224 $\mu$m and  2.248 $\mu$m 
lines  were not detected through all the slitlets, or have lower SNR (by a
factor $\sim$~5 and $\sim$~10, respectively) than that of the 
2.122 $\mu$m line. The Q-branch  lines  are 
brighter, but they lie at the edge of the spectral range covered with the
spectral configuration  used (at 2.4 $\mu$m), and the accuracy reached in the 
wavelength calibration was lower (rms of 0.40 \AA, $\sim$~5.7 \kms) than that
 achieved for  the 2.122 $\mu$m line emission (rms of 0.25 \AA,
$\sim$~4 \kms).

\begin{figure*}
\includegraphics[angle=-270,width=160mm,clip]{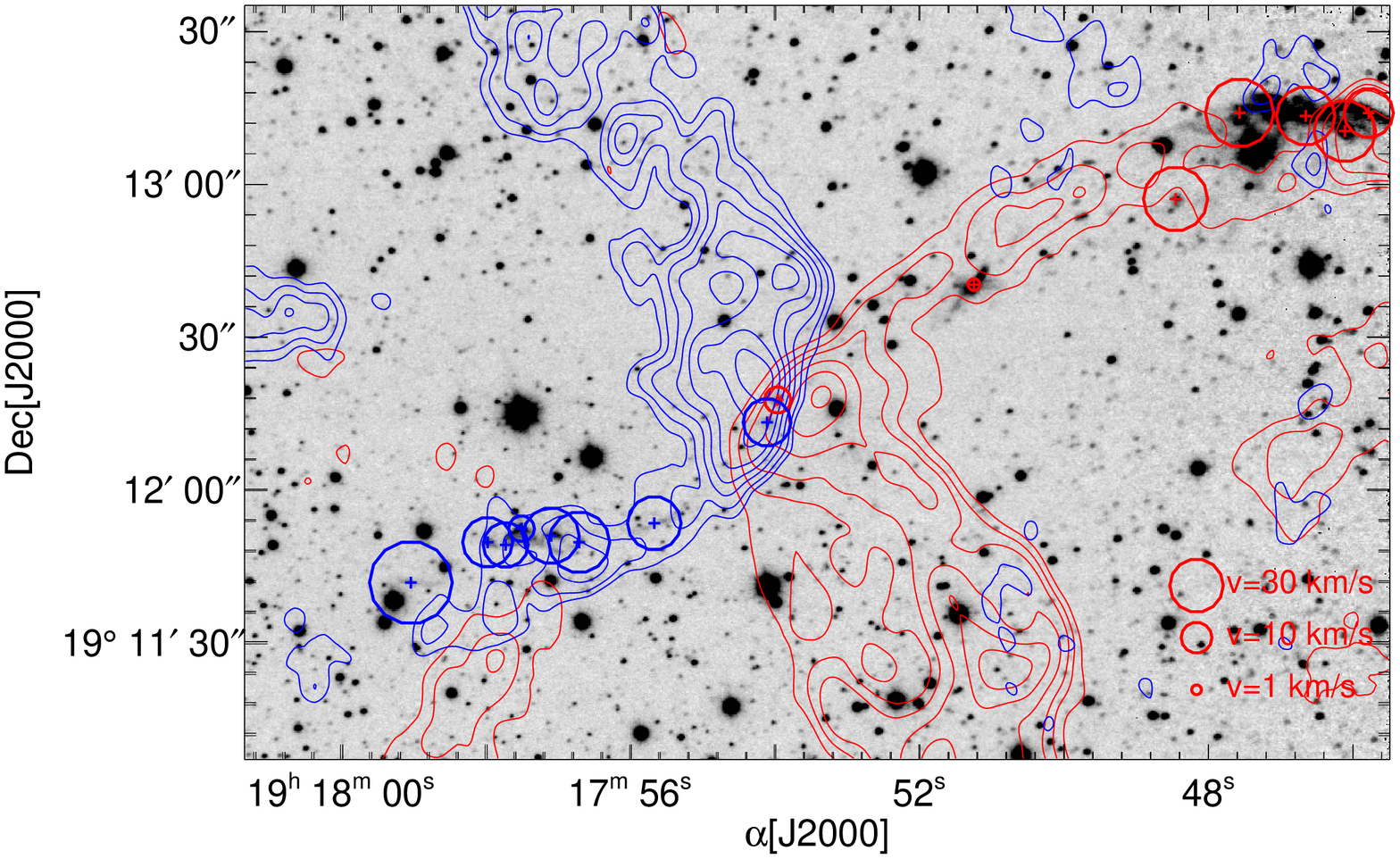}
\caption{H$_2$ image of the L723 field (gray scale) supperposed to the contours of the redshifted and blueshifted CO
outflows from \citet{Lee02}. The red and blue open circles of different sizes represent the
radial velocity obtained from the 2.122~$\mu$m line at the knot positions marked by the crosses.
(The scale of velocity is shown at the botton right corner).
\label{vlsrimage}} 
\end{figure*}

\subsubsection{Kinematics of the H$_2$ emission}

In order to search for a kinematic trend
along the HH~223 outflow, we derived the  
radial velocity\footnote{All the velocities in the paper are given with respect to the 
velocity adopted for the parent cloud,  \vlsr=+10.9~\kms (\citealp{tor86})} from the  
spectra  obtained by averaging the
signal within the full aperture of each slitlet (Table \ref{vcenslit} and  
Fig.~\ref{censlitav}).  The velocities derived in this way  correspond to the mean
velocity
within the aperture,  and do not account for  velocity shifts at spatial
scales smaller than the length of the corresponding slitlet, which  in some cases 
covers several knots (see Fig.~\ref{mapah2}).

\begin{figure}
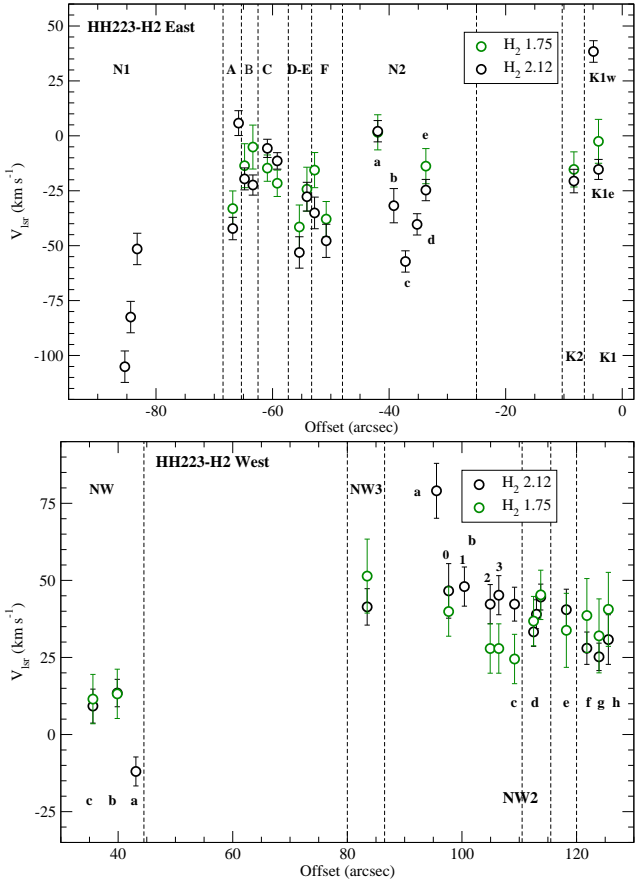

\includegraphics[width=84mm,clip]{fig9a.eps}
\includegraphics[width=84mm,clip]{fig9b.eps}
\caption{Radial velocities of the identified knots through the slitlets 
as a
function of the distance to SMA~2. Dashed vertical lines are plotted to separate the emission coming
from different slitlets.
\label{vknots}}
 \end{figure}


\begin{figure}
\includegraphics[width=84mm,clip]{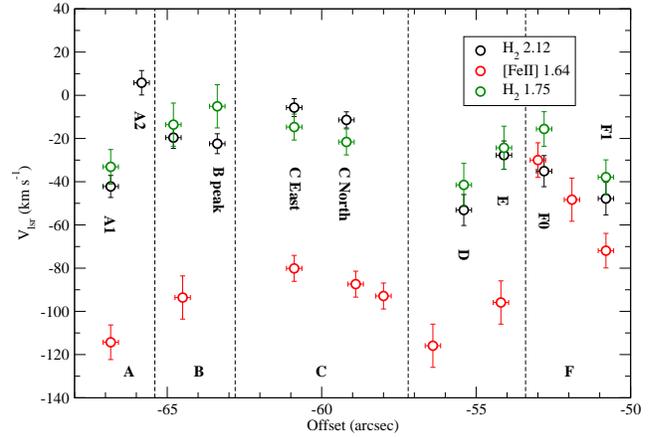}
\caption{Close-up of Fig. \ref{vknots} showing the radial 
velocities of the identified knots through the slitlets covering HH223A-F (the near-infrared
counterpart of the HH~223 linear emission feature, detected in H$\alpha$ and \sii\ lines).
\label{main}} 
\end{figure}


\begin{table*}
\begin{minipage}{140mm}
\caption{Radial velocities (\vlsr)$^{1}$ of the H$_2$ 2.122 $\mu$m emission for the knots
identified within the slitlets}
\begin{tabular}{llrrr}
\hline
Slitlet&Knot    &Length$^{2}$ &Offset $^{3}$&  \vlsr $^{4}$\\
       &HH~223- &(arcsec)& (arcsec)& (\kms)\\
\hline
s1     &N1      &      &	  &               \\
       &N1a     & 1.00 &$-85.4$   &$ -105.1\pm14.3$\\
       &N1b     & 0.75 &$-84.4$   &$  -82.5\pm14.3$\\
       &N1c     & 1.50 &$-83.2$   &$  -51.5\pm14.3$\\
s2     &A       &      &	  & 	        \\
       &A1      & 1.00 &$-66.8$   &$  -42.2\pm10.2$\\
       &A2      & 0.75 &$-65.8$   &$   +5.8\pm11.3$\\
s3     &B       &      &	  &		     \\
       &B-East  & 1.00 &$-65.1$   &$   -19.6\pm10.0$\\
       &B-peak   & 1.00 &$-63.4$   &$   -22.4\pm\phantom{0} 9.2$\\
s4     &C       &      &	  &		       \\
       &C-East  & 1.00 &$-60.9$   &$	  -5.7\pm\phantom{0} 8.3$\\
       &C-North & 1.50 &$-59.0$   &$	 -11.4\pm\phantom{0} 7.5$\\
s5     &D-E     &      &          &		      \\
       &D       & 1.00 &$-55.5$   &$   -53.1\pm14.3$\\
       &E       & 1.50 &$-54.2$   &$   -27.7\pm13.1$\\
s6     &F       &      &	  &		       \\
       &F0      & 1.50 &$-52.7$   &$   -35.1\pm14.4$\\
       &F1      & 1.75 &$-50.9$   &$   -47.8\pm15.1$\\
s7     &N2      &      &	  &		     \\
       &N2a     & 2.00 &$-42.0$   &$	+2.1\pm\phantom{0} 9.7$\\
       &N2b     & 1.75 &$-39.2$   &$   -31.8\pm15.6$\\
       &N2c     & 2.00 &$-37.2$   &$   -57.2\pm\phantom{0} 9.7$\\
       &N2d     & 1.75 &$-35.2$   &$   -40.3\pm\phantom{0} 9.7$\\
       &N2e     & 1.00 &$-33.7$   &$   -24.7\pm\phantom{0} 9.7$\\
s8     &K2      &      &          &		     \\
       &K2-peak  & 1.00 &$-8.3$    &$   -20.6\pm10.7$\\
s9     &K1      &      &          &		    \\
       &K1-East & 1.25 &$-4.9$    &$   +38.4\pm\phantom{0} 9.8$\\
       &K1-West & 1.00 &$-4.1$    &$   -15.3\pm\phantom{0} 9.1$\\
s11    &H2-NW   &      &	  &		    \\
       &H2-NWc  & 2.75 &$35.6$    &$  +9.2 \pm 11.0$\\
       &H2-NWb  & 2.75 &$39.8$    &$  +13.5\pm\phantom{0} 8.9$\\
       &H2-NWa  & 3.0  &$43.1$    &$  -12.0\pm\phantom{0} 9.4$\\
s12    &H2-NW3  &      &    	  &	     \\
       &H2-NW3pe& 1.75 &$83.5$    &$  +41.4\pm11.8$\\
s13    &H2-NW2  &      &	  &		   \\
       &H2-NW2a & 1.00 &$ 95.6$   &$  +79.1\pm17.7$\\
       &H2-NW2b0& 1.25 &$ 97.7$   &$  +46.6\pm17.7$\\
       &H2-NW2b1& 1.75 &$100.4$   &$  +48.0\pm12.7$\\
       &H2-NW2b2& 1.25 &$104.9$   &$  +42.3\pm12.7$\\
       &H2-NW2b3& 2.00 &$106.4$   &$  +45.2\pm12.7$\\
       &H2-NW2c & 2.00 &$109.2$   &$  +42.3\pm11.0$\\
s14    &H2-NW2  &      &	  &		    \\
       &H2-NW2d1& 0.75 &$112.5$   &$  +33.3\pm\phantom{0} 9.4$\\
       &H2-NW2d2& 0.75 &$113.0$   &$  +39.0\pm\phantom{0} 9.4$\\
       &H2-NW2d3& 1.00 &$113.8$   &$  +44.6\pm\phantom{0} 8.4$\\
s15    &H2-NW2  &      &	  &		    \\
       &H2-NW2e & 2.25 &$118.2$   &$  +40.5\pm13.3$\\
s16    &H2-NW2  &      &	  &		    \\
       &H2-NW2f & 1.75 &$121.8$   &$  +28.0\pm10.5$\\
       &H2-NW2g & 1.50 &$123.9$   &$  +25.2\pm\phantom{0} 8.9$\\
       &H2-NW2h & 1.00 &$125.5$   &$  +30.8\pm16.1$\\
\hline 
\end{tabular}
\label{vrknots}
\begin{list}{}
\item
$^{1}$ From Gaussian fits to the H$_2$ 2.122 $\mu$m line of the spectra 
averaged within apertures of 1~arcsec width and lengths given in column 3.\\ 
$^{2}$ Length along the slitlet of 1 arcsec width, including emission 
from the knot.\\
$^{3}$ Offsets from the position of SMA~2 to the centre of the apertures. 
Typical values of the error are $\leq$ 0.5~arcsec\\
$^{4}$ Errors, derived as in Table \ref{vcenslit}.\\
\end{list}
\end{minipage}
\end{table*} 

The velocities derived in this way are given in 
Table \ref{vcenslit} and  Fig. \ref{censlitav} and show a bipolar pattern: the velocity
along the outflow changes from
negative (blueshifted) in the southwest to positive (redshifted)  
to the northwest. 
We found a mean velocity of --32.8 \kms\ for 
the knots observed  at distances ranging from 40 to 85~arcsec
to the southeast of SMA~2 (the radio-continuum source that hides the outflow exciting
source).
A mean velocity of +32.5 \kms\ is found 
for the 
knots located at distances ranging from 40 to 130~arcsec to the northwest of 
SMA~2. 
The velocity changes its sign (from negative to positive values) in the neighbourhood of 
the HH~223-K1 nebula, \ie\ close to the position of SMA~2. 

It is worth noting that the  HH 223 H$_2$ outflow lies projected onto the pair of
lobes of one of the two bipolar CO outflows (the larger, east-west one) detected
in L723. As a general trend, the H$_2$ knots with blueshifted velocities are
projected onto the blueshifted CO outflow lobe, while the H$_2$ knots with
redshifted velocities are projected onto the redshifted CO outflow lobe, as can 
be easily visualized in Fig.~\ref{vlsrimage}. 
This spatial 
coincidence between the CO and H$_2$ velocity signs 
gives support to the existence of a physical relationship between the H$_2$ and CO
outflows.

Furthermore, we searched for changes in velocity at spatial
scales smaller than the length of the slitlets.  In Fig.~\ref{mapah2} (black contours), 
we plot position-velocity
(PV) maps of the H$_2$~2.122 $\mu$m  emission through all the slitlets. 
From these maps, and comparing
with the H$_2$ {\it K} narrow-band continuum-subtracted image of the 
field (see \citealp{Lop10}), we identified which knots were intersected by
each slitlet. Then,  we obtained the one-dimensional spectrum of 
each of these knots by
averaging the signal over the length of the slitlet encompasing the corresponding
knot (\ie\ within windows 1 arcsec wide and  lengths given in Table
\ref{vrknots}, column 3), and derived the velocity of the knots 
(Table \ref{vrknots}, column 5). The distance, measured as
the offset from the position of SMA~2 to  the emission peak of the knot, 
is also given in
Table \ref{vrknots}.  The velocities obtained for the knots
projected onto the blue/red lobes of the east-west CO outflow  are plotted in Fig.
\ref{vknots} (black dots). These radial velocities will be used later to
derive the full spatial velocity field of the near-infrared HH~223
outflow. At small scales 
we found
a rather complex kinematics within the extended nebular emission features,
which confirms the previous IFS results (\citealp{Lop12}) for the HH~223 
optical counterpart.
Note in addition that we  detected some additional
H$_2$ features from the MOS spectra that were not detected in the previous
H$_2$ narrow-band image. 
 
Finally, and in order to get further evidence of the kinematic pattern
found from the H$_2$~2.122~$\mu$m line, we obtained the velocity of the
knots from the  1.748 $\mu$m line, the brightest H$_2$
line detected in  most of the {\it H}-band spectra. The results are shown
in Figs.~\ref{vknots} and \ref{main} (green dots). As can be seen from the
figures, they are in good agreement with the results
found from the H$_2$ {\it K}-band  line.
Concerning the values of the radial velocities, they are  consistent with those
derived from the 2.122~$\mu$m line. Concerning the spatial brightness distribution, the
knot structures found from the 2.122~$\mu$m line are also identified in the {\it H}-band
maps, and the positions
of the peaks of the two H$_2$ emission lines are  coincident. However, the general trend
found in all the condensations is that  the emission of the 1.748~$\mu$m line is
detected within a narrower   velocity range (by $\sim$ 40--60~\kms) than the emission
from the 2.122~$\mu$m line (at the same SNR levels).

\subsubsection{Kinematics of the ionized outflow gas}


\begin{table}
\caption{Radial velocities (\vlsr)$^{1}$ of the \fe\ 1.644 $\mu$m emission for the knots
identified within the slitlets}
\begin{tabular}{llrrr}
\hline
Slitlet&Knot    &Length$^{2}$ &Offset $^{3}$&  \vlsr $^{4}$\\
       &HH~223- &(arcsec)& (arcsec)& (\kms)\\
\hline

s2     &A       &      &	  & 	        \\
       &A1      & 2.75 &$-66.8$   &$  -114.3\pm10.8$\\
s3     &B       &      &	  &		     \\
       &B-peak   & 1.00 &$-64.5$   &$   -93.6\pm\phantom{0} 9.6$\\
s4     &C       &      &	  &		       \\
       &C-East  & 1.25 &$-60.9$   &$	 -80.1\pm 11.3$\\
       &C-North & 1.50 &$-58.9$   &$	 -87.4\pm\phantom{0} 9.5$\\
       &C-peak   & 2.25 &$-58.0$   &$	 -92.9\pm\phantom{0} 8.5$\\
s5     &D-E     &      &          &		      \\
       &D       & 1.25 &$-56.4$   &$   -115.9\pm 13.1$\\
       &E       & 2.00 &$-54.2$   &$   -95.9\pm 12.3$\\
s6     &F       &      &	  &		       \\
       &F0      & 1.25 &$-53.0$   &$   -30.0\pm 13.6$\\
       &F1a     & 0.75 &$-51.9$   &$   -48.3\pm 15.1$\\
       &F1b     & 2.25 &$-50.8$   &$   -71.9\pm 14.1$\\
\hline 
\end{tabular}
\label{vrfeknots}
\begin{list}{}
\item
$^{1}$ From Gaussian fits to the \fe\ 1.644 $\mu$m line of the spectra 
averaged within apertures of 1~arcsec width and lengths given in column 3.\\ 
$^{2}$ Length along the slitlet of one arcsec width, including emission 
from the knot.\\
$^{3}$ Offsets from the position of SMA~2 to the centre of the apertures. 
Typical values of the error is $\leq$ 0.5~arcsec\\
$^{4}$ Errors, derived as in Table \ref{vcenslit}.\\
\end{list}
\end{table} 

We derived the kinematics of the  ionized gas by obtaining the velocity from  the
brightest \fe\ lines in the {\it J}  and {\it H} bands  (\ie\ the 1.257~$\mu$m and
1.644~$\mu$m lines, respectively).  The results found are given in Table~\ref{vrfeknots}
and are drawn in Fig.~\ref{main} (red dots). The \fe\ \vlsr\ values appear blueshifted,
ranging from $\sim$~--30 to $\sim$~--115 \kms.  These velocities show that the  ionized
gas of the outflow (traced by the \fe\ emission) is more  blueshifted than  the molecular
gas (traced by the H$_2$ emission).

\subsubsection{Individual outflow features}

In the following, we will briefly discuss the kinematic structure found 
at small scales, by examining each of the 
nebulae sampled with MOS through all the slitlets.

{\bf HH~223-A to HH~223-F}\\
Five slitlets (S2 to S6, from east to west) were positioned  covering the
knots HH~223-A to F, the knotty, undulating emission of $\sim$ 22 arcsec
in length, at $\sim$ 1 arcmin southeast of SMA~2. Note
that this structure  is  the optical counterpart of the H$\alpha$ ``linear
emission feature'' reported by \citet{Vrb86}, and corresponds to the
Herbig-Haro object 223.

Figure~\ref{mapah2} (left column) displays the PV maps of the H$_2$
emission lines at 2.122~$\mu$m (black) and 1.748~$\mu$m (green), and \fe\
at 1.644~$\mu$m (red) through the S2 to S6 slitlets. The knots intersected
by the slitlets, and  identified  from  the H$_2$ {\it K}-narrow-band,
continuum substracted image of the field, have  been  labeled accordingly.
Figure~\ref{main} displays a close-up of  Fig.~\ref{vknots} showing with 
more detail  the behaviour of the velocity as a function of the
angular distance to SMA~2. The spectrum of each knot was obtained by
averaging the signal within the apertures quoted in Table~\ref{vrknots}, and
the velocities were derived from the 1.748~$\mu$m and
2.122~$\mu$m H$_2$, and 1.644~$\mu$m \fe\ lines. Both emissions, the
neutral (traced by H$_2$) and the ionized (traced by
\fe) appear blueshifted. As expected, the velocities derived from \fe\
are more blueshifted  than those derived for both
H$_2$ lines, which in turn are consistent between them. Accordingly with
what was found in the PV maps  (see Fig.~\ref{mapah2}) and in the {\it H}-
and {\it K}-narrow-band images (\citealp{Lop10}), there is some
shift between the spatial distribution of the \fe\ and H$_2$
emissions:  offset values $\geq$~1 arcsec are found between the peaks of
the \fe\ and H$_2$ emissions in knots HH~223-B, -C (that corresponds to -C
North in H$_2$),  and -D. Furthermore, there are some substructures that
were detected in H$_2$ but did not appear in \fe\ (\eg\ HH~223-A2) while 
some \fe\ substructures were not detected in H$_2$ (\eg\ a
third condensation lying between HH~223-F0 and -F1). As it was noted,
these facts may be indicative of shocks of different degree of excitation
coexisting along the HH~223-A to -F emission region.

A trend for the velocity can be found, consisiting in an increase of
the absolute value of the velocity, suggestive of an acceleration,  
from east to west, towards the location of SMA~2.  
The general trend seems to be broken at a few positions
(\eg\ at A1 and D knots).  However, this might be expected because of the
highly complex kinematics of HH~223-A-F already found at optical wavelengths,
where a fully spatial sampling of the emission was obtained with Integral
Field Spectroscopy (\citealp{Lop12}). One indication of such a highly
complex kinematics can be found by examining the PV maps of the 2.122
$\mu$m line  of HH~223-A-F at spatial scales of $\sim$ 1 arcsec:
as can be seen from Fig. \ref{mapah2}, a  change in velocity
can  be appreciated along HH~223-A (\ie\ in S2). Moving from east to west
through the slitlet, we found an intensity peak of  emission, 
labeled as knot HH~223-A1 in Table~ \ref{vrknots}, and another  secondary
emission enhancement (HH~223-A2 in Table~ \ref{vrknots}), their peaks
being separated $\sim$ 1 arcsec.  From the spectra obtained for these two 
HH~223-A knots
we found a difference in velocity of $\sim$~50~\kms\ between them, 
being the eastern knot more blueshifted.  It is worth
noting that HH~223-A  presents two emission peaks separated $\sim$ 1
arcsec, partially resolved in the H$\alpha$ images (\citealp{Lop09}).
The radial velocities derived from the H$\alpha$ and \sii\  lines in 
long-slit spectra through HH~223-A show the same behaviour
as in the case of the H$_2$ line: a change in velocity of
$\sim$~60~\kms\ between these two condensations of HH~223-A, being  the
eastern condensation (HH~223-A1) the most blueshifted. 
Furthermore, the velocity trend shown by the 2.122~$\mu$m line 
through HH~223-A is also  found in  other H$_2$  lines (\eg\ the
2.224~$\mu$m and the  2.406~$\mu$m). In
contrast, the two H$_2$ substructures found within 
HH~223-A were not identified in
the PV map of the  \fe\ 1.644~$\mu$m line. The position of the \fe\ emission
peak coincides  with HH~223-A1, while no clear \fe\ counterpart
is found for the other H$_2$ substructure (HH~223-A2) . Hence, we
confirm that there are two unresolved clumps  within HH~223-A, with 
different excitation conditions, which may be  
originated in shocks with
different strengths.

In spite of the lack of detection of H$_2$ emission  from the knot HH~223-B
in the narrow-band images, 
we were able to detect H$_2$ emission through the
slitlet positioned  on this knot (S3).
From the H$_2$ 2.122~$\mu$m line, we derived
the velocity at the intensity peak of the emission, and at a secondary
weaker substructure lying to its southeast  (named HH~223-B-peak and
-B-east, respectively; see Table \ref{vrknots}). At both positions the
emission appears blueshifted, although no significant variation in
velocity along the knot is found. In addition, we detected emission from
other H$_2$ lines, both in the {\it H} and {\it K} bands, the behaviour being
consistent with that of the 2.122~$\mu$m emission (see \eg\ the PV map of the
1.748~$\mu$m emission in Fig.~\ref{mapah2}). Note, in contrast, the
spatial displacement found between the \fe\ and H$_2$ emissions, inferred
from their PV maps (Fig.\ref{vrknots}). The position of the \fe\ emission
peak is  coincident (within 0.2 arcsec) with the peak of
HH~223-B-east, the H$_2$ weaker substructure, but it is offset by 1.4
arcsec from HH~223-B-peak.

 The slitlet S4 included emission from the intensity peak of the
 bow-shaped feature labeled HH~223-C in the narrow-band H$_2$ image,
 and from a weaker condensation, southeast of it (labeled HH~223-C-North and
 C-East, respectively, in this work). A spatial shift between the 
 \fe\ and H$_2$ emissions was already found in the {\it H} and {\it K}
  narrow-band images, and also can be appreciated in the PV maps of this
 work. The peak of the \fe\ emission is  offset by $\sim$~1.2 arcsec west
 from the peak of the H$_2$ emission (HH~223-C-North). 

 The slitlet S5 included emission coming from the bow-shaped feature HH~223-E, and
from HH~223-D. We found an offset of $\sim$~1 arcsec between the peaks of the 
\fe\ and H$_2$
emissions in HH~223-D, while there is not a significant shift between the peaks of these emissions
 for HH~223-E. Finally, 
 S6 included the emission from the filamentary feature HH~223-F. 
 The spatial brightness distribution
of the ionized (\fe) and neutral (H$_2$) emissions are in general  coincident, although
some 
differences are found. Two condensations, F0 and F1,  are identified in the PV 
map of both, H$_2$ and \fe\
emissions,  
while the \fe\ emission presents a third  condensation located between them. 
The trend of decreasing velocity from east to west is seen through HH~223-F for both
ionized and neutral emissions.

{\bf HH~223-N2}\\
The slitlet S7 was positioned along the H$_2$ faint filamentary
structure $\sim$~30 arcsec northwest of HH~223-A. We identified several
brightness enhancements that were detected only in the brighter H$_2$ 
lines (\eg\ 1.748~$\mu$m and 2.122~$\mu$m). In contrast, emission from
\fe\ was not detected in the spectral range observed. No clear trend along
the filament for the radial velocity  was found. On the other hand, the
tangential velocity cannot be derived from  the narrow-band images due to
the low contrast brightness distribution of the clumps. Hence, it is not  possible to
infer whether the observed kinematics could be tracing some changes in the
outflow orientation at this region.

{\bf HH~223-K2, --K1}\\
Two slitlets were  positioned on the H$_2$ nebulae closest to the location of SMA~2. The
H$_2$ emission from HH~223-K2 (through S8) appears blueshifted, without significant
velocity variations along the region mapped.  In contrast, the velocity derived along S9 
shows some variation from east (blueshifted) to west (redshifted). Note that S9 was
positioned closely perpendicular to HH~223-K1. Hence, the K1-West emission peak is  the
closest emission  to the wall of the cavity opened by the CO outflow. In spite of these
slitlets being the closest  to the  YSO position,  we were not able to detect emission from
\fe\ lines.

{\bf HH~223-H2-NW}\\
Slitlet S11 sampled the emission along the filamentary feature
HH~223-H2-NW, crossing the two bright knots (labeled NWa and NWb, from
west to east), and the weaker, diffuse emission labeled NWc. From the
2.122 $\mu$m line, we derived similar redshifted velocity values for
knots NWc and NWb, as  expected for the H$_2$ knots lying projected onto
the redshifted CO outflow lobe. In contrast, a blueshifted velocity was
derived at NWa. At the brighter knots (NWa and NWb), emission from other 
H$_2$ lines of the {\it K}-band
was also detected. Thus, in addition
to the 2.122 $\mu$m line, we used the 2.224 $\mu$m and 2.248 $\mu$m lines
for deriving their velocities. From  
these three lines,  we found mean  velocities   of
+13.3 $\pm$~0.2 \kms\ and --12.0 $\pm$~0.9 \kms\ for knots NWb and NWa
respectively. Then, we concluded that this difference in velocity between
knots a and b, c is reliable.

{\bf HH~223-H2-NW3}\\
Slitlet S12 was  positioned on this isolated, compact ($\sim$~2~arcsec in diameter)
emission  feature, seen in the narrow-band images towards the southeast of the filamentary
feature HH~223-H2-NW2. We were not able to identify substructures  within the knot.   The
velocity of the knot is redshifted, without significant variations along it. 

{\bf HH~223-H2-NW2}\\
This bright H$_2$ filamentary feature is seen towards the northwestern
edge of the observed field. Part of it coincides with the {\it K'} linear nebula
reported by \cite{Hod94}, and has a faint, less extended optical
counterpart visible in the H$\alpha$ line (\citealp{Lop06}). The filament
appeared resolved in several brightness enhancements, labeled {\it a} to {\it g}, 
in the narrow-band H$_2$ images (\citealp{Lop10}).

Up to four slitlets (S13 to S16) were positioned along this filament in
order to cover the emission coming from most of its bright knots. S13
included emission   from a part of the faint,
more extended nebula NW2a, the eastern part of the filament, from NW2b, which was
resolved into several smaller-scale emission enhancements (labeled b0 to
b3, from east to west along the filament), and most of the NW2c knot. S14 included
emission from NW2d, which showed  three emission enhancements (labeled d1
to d3, from east to west along the feature). S15 crossed NW2e, which
showed a compact,  single-peaked emission in all the detected H$_2$ lines.
Finally, S16 included emission coming from NW2f, NW2g, and from NW2h, a 
more difuse, fainter  emission,  first identified from MOS spectra. 

Using the 2.122 $\mu$m line, we derived the velocity for each of the
condensations identified in the MOS spectra. The general trend shows
a slight decrease of the redshifted velocity,  which
diminished by a factor of $\sim$ 2 in $\sim$  25 arcsec 
along  the filament, from east to west.

\subsection{Proper motions}


\begin{figure*}
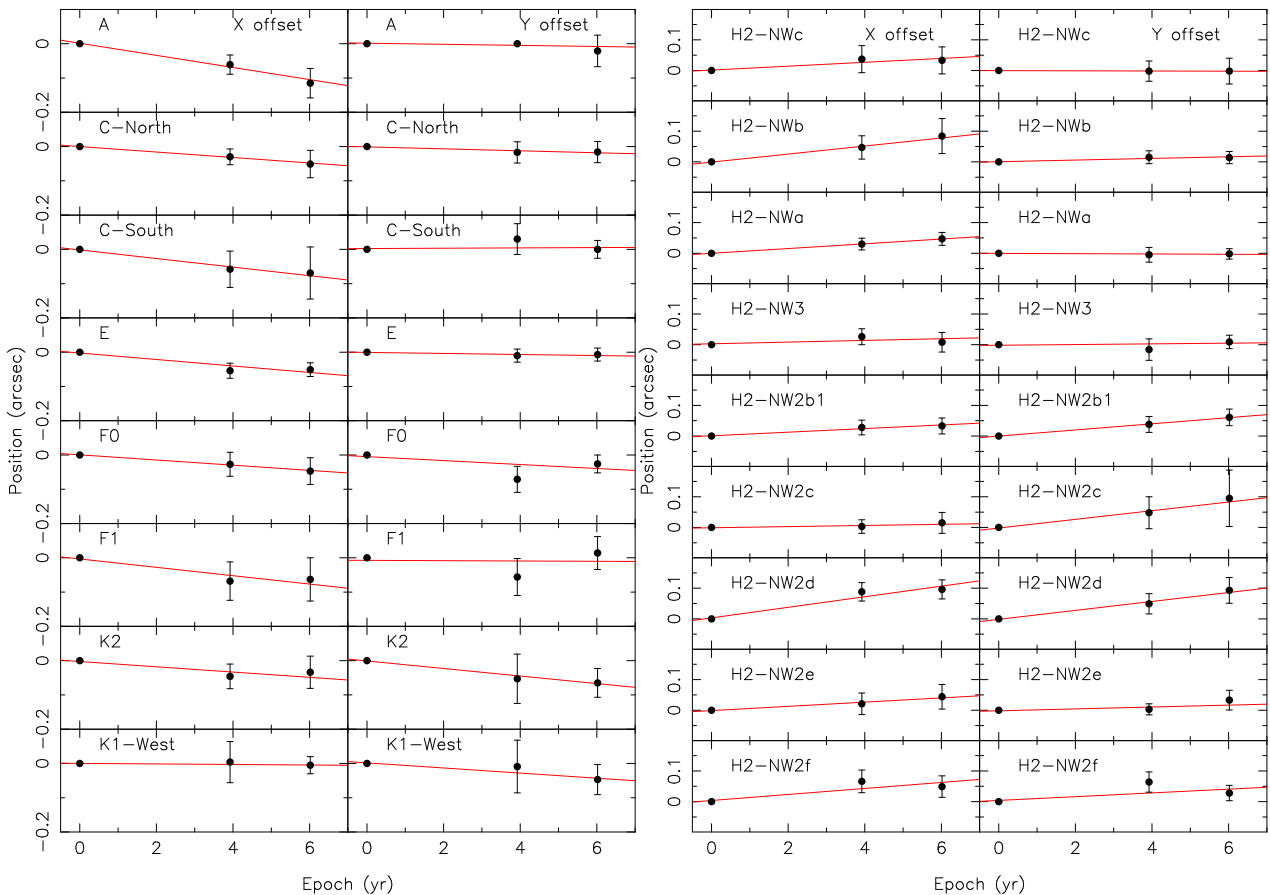

\begin{minipage}{168mm}
\includegraphics[width=84mm,clip]{fig11a.eps}%
\includegraphics[width=84mm,clip]{fig11b.eps}
\caption{Proper motions derived for the HH~223 knots labeleled in the panels.
For each epoch, a circle represents the displacements (in arcsec) in the
$x$-direction (left) and $y$-direction (right) of the corresponding knot
measured from the first-epoch image, which defines the origin of the time-scale
(set on 2006 July 20). Errors are indicated by vertical bars. The least-squared
fits derived for the displacements of each knot are shown by the red lines. The
proper motion of the knots in the $x$ and $y$ direction are derived from the
slopes of these lines.
\label{ofxy}}
\end{minipage} 
\end{figure*}

\begin{figure*}
\includegraphics[angle=-90,width=160mm,clip]{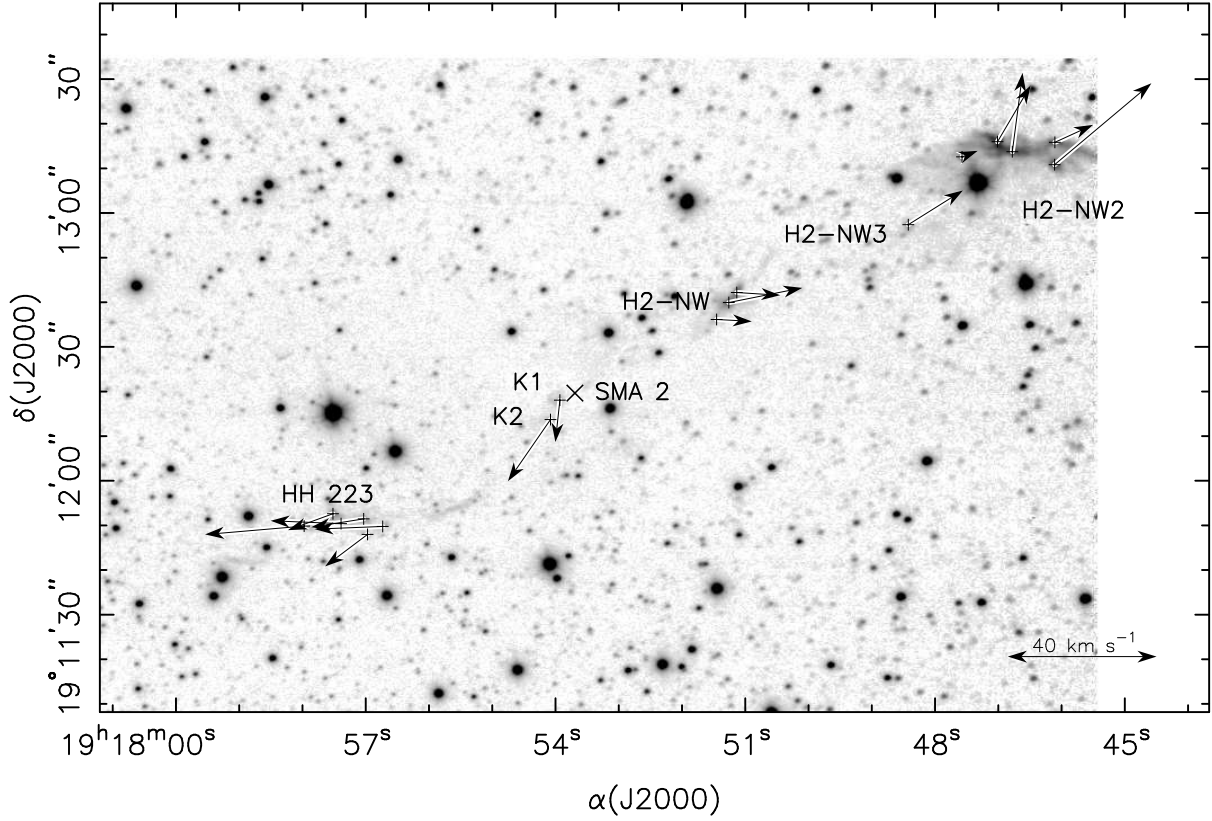}
\caption{Image of the L723 field through the H$_2$ 2.122 $\mu$m line filter
(continuum is not subtracted). The arrows indicate the proper motion velocity
(\vtan) derived for each knot. The position of SMA~2 is indicated by a $\times$ sign. 
The scale of the arrows (in \kms) is indicated
by  the double headed arrow at the bottom right corner of the map. 
\label{mp}} 
\end{figure*}


\begin{table*}
\begin{minipage}{160mm}
\caption{\label{t_mp} Positions and Proper Motions of H$_2$ Knots}
\begin{tabular}{lrrrrrrrr}
\hline
Knot
&
${x}^a$&  
${y}^a$&  
${\mu_x}^b$&
${\mu_y}^b$&
${\epsilon_x}^c$&
${\epsilon_y}^c$&
{\vtan}$^d$&
PA
\\
HH~223-&
(arcsec)& 
(arcsec)& 
(arcsec yr$^{-1}$)&
(arcsec yr$^{-1}$)&
(arcsec)&
(arcsec)&
(km s$^{-1}$)&
(deg)
\\
\hline
 A           & $-60.6$& $-29.8$& $-0.0177\pm0.0014$& $-0.0015\pm0.0015$& $0.0059$& $0.0057$& $25.3\pm2.0$& $ 95\pm\phantom{0}5$\\
 C-North     & $-54.2$& $-27.0$& $-0.0080\pm0.0003$& $-0.0029\pm0.0006$& $0.0012$& $0.0027$& $12.2\pm0.5$& $110\pm\phantom{0}4$\\
 C-South     & $-52.4$& $-29.2$& $-0.0126\pm0.0013$& $ 0.0005\pm0.0020$& $0.0053$& $0.0097$& $17.9\pm1.8$& $ 88\pm\phantom{0}9$\\
 E           & $-47.3$& $-28.2$& $-0.0094\pm0.0018$& $-0.0015\pm0.0005$& $0.0081$& $0.0022$& $13.5\pm2.5$& $ 99\pm\phantom{0}3$\\
 F0          & $-46.4$& $-31.6$& $-0.0076\pm0.0003$& $-0.0057\pm0.0040$& $0.0014$& $0.0192$& $13.5\pm3.5$& $127\pm   19$\\
 F1          & $-43.0$& $-29.9$& $-0.0122\pm0.0026$& $-0.0005\pm0.0057$& $0.0114$& $0.0261$& $17.4\pm3.7$& $ 92\pm   27$\\
 K2          & $ -5.5$& $ -5.9$& $-0.0076\pm0.0024$& $-0.0110\pm0.0007$& $0.0100$& $0.0036$& $19.0\pm2.1$& $145\pm\phantom{0}8$\\
 K1-West     & $ -3.4$& $ -1.6$& $-0.0008\pm0.0004$& $-0.0075\pm0.0015$& $0.0020$& $0.0071$& $10.7\pm2.1$& $174\pm\phantom{0}3$\\
 H2-NWc   & $ 31.7$& $ 16.5$& $ 0.0063\pm0.0014$& $-0.0003\pm0.0001$& $0.0062$& $0.0002$& $ 9.0\pm1.9$& $-93\pm\phantom{0}1$\\
 H2-NWb   & $ 34.5$& $ 20.3$& $ 0.0131\pm0.0008$& $ 0.0026\pm0.0005$& $0.0032$& $0.0023$& $19.1\pm1.1$& $-79\pm\phantom{0}2$\\
 H2-NWa   & $ 36.3$& $ 22.5$& $ 0.0078\pm0.0001$& $-0.0005\pm0.0003$& $0.0001$& $0.0012$& $11.1\pm0.1$& $-94\pm\phantom{0}2$\\
 H2-NW3   & $ 74.7$& $ 37.7$& $ 0.0098\pm0.0031$& $ 0.0061\pm0.0036$& $0.0140$& $0.0171$& $16.4\pm4.6$& $-58\pm   17$\\
 H2-NW2b1 & $ 86.7$& $ 52.9$& $ 0.0028\pm0.0021$& $ 0.0011\pm0.0016$& $0.0089$& $0.0076$& $ 4.3\pm2.8$& $-69\pm   31$\\
 H2-NW2c  & $ 94.7$& $ 56.3$& $ 0.0059\pm0.0006$& $ 0.0101\pm0.0002$& $0.0028$& $0.0008$& $16.7\pm0.5$& $-31\pm\phantom{0}3$\\
 H2-NW2d  & $ 98.0$& $ 54.1$& $ 0.0018\pm0.0007$& $ 0.0141\pm0.0014$& $0.0028$& $0.0056$& $20.2\pm2.0$& $ -7\pm\phantom{0}3$\\
 H2-NW2e  & $107.5$& $ 51.2$& $ 0.0173\pm0.0023$& $ 0.0145\pm0.0011$& $0.0104$& $0.0048$& $32.1\pm2.7$& $-50\pm\phantom{0}4$\\
 H2-NW2f  & $107.5$& $ 56.2$& $ 0.0068\pm0.0007$& $ 0.0031\pm0.0019$& $0.0032$& $0.0074$& $10.7\pm1.5$& $-65\pm   13$\\
\hline
\end{tabular}

\medskip
$^a$
{Offsets from the position of SMA 2, $\alpha(J2000)=19^\mathrm{h}17^\mathrm{m}53\fs694$, $\delta(J2000)=19^\circ12'19\farcs68$,
of the intensity peak in the 2006 image. For the more extended knots, Cs and H2NW2f+g, the
position of the center of the box used for the proper motion determination is given. 
The $x$ axis points westwards, and the $y$ axis northwards.}
$^b$
{Error values are the error in the proper motion fit to the three epochs observed.}
$^c$
{rms residual of the knot positions in the proper motion fit, 
$\epsilon=\sigma \sqrt{1-r^2}$, where $\sigma$ is the
standard deviation and $r$ is the correlation coefficient.}
$^d$
{Proper motion velocity, assuming a distance of 300 pc.}
\end{minipage}
\end{table*}
The proper motions of the knots along the HH~223 outflow were derived from
three H$_2$ narrow-band  images of the L723 field obtained at three
differents epochs,  spanning a total of 6 years (see Table~\ref{ta:obslog}).

The three images were converted into a common reference system. The
positions of  ten field stars, common to all the frames, were used to
register the images. The {\it geomap} and  {\it geotran} 
tasks of {\small IRAF} were applied to
perform a linear transformation, with six free parameters that take into
account translation, rotation and magnification between different frames.
After the transformation, the typical rms of the difference in position
for the reference stars between the reference epoch and the other two
epochs was $\sim$~0.03~arcsec in both coordinates. The pixel size was
found to be 0.252~arcsec.

We defined boxes that included the emission of the individual
condensations in each epoch. The position offset of the second and third
epochs with respect to the first epoch (taken as reference) was estimated
by cross-correlation (see the description of this method in
\citealp{Rei92};  \citealp{Lop96}).  The uncertainty in the position of
the correlation peak was estimated in the same way  as was done by
\citet{Ang07}, through the scatter of the correlation peak positions
obtained from boxes differing from the nominal one in $\pm2$ pixels.   The
error adopted for each coordinate for the offset for epoch $i$,
$\epsilon_i$, was twice the uncertainty in the correlation peak position,
added quadratically to the rms alignment error.

In Fig.\ \ref{ofxy}  we show the position offsets in $x$ and $y$ 
measured for the knots. 
The proper motions in the $x$ and $y$ direction, $\mu_x$ and $\mu_y$,  were
obtained for each knot as the slope of the regression lines fitted to the offset
positions for the three epochs (see Fig.\ ~\ref{ofxy}).
The proper motions obtained are shown
in Table \ref{t_mp} and in Fig.\ \ref{mp}.
The errors assigned to the proper motions, in both $x$ and $y$, are the 
formal errors of the slope of the linear regression fits for each knot. 
The errors appear to be rather small as a consequence of the wide time  span 
used for calculating the proper motions. The quality of the linear regression
fits for each knot is indicated by the small values of the residuals in $x$,
$\epsilon_x$, and $y$, $\epsilon_y$, (see Table \ref{t_mp}).

As can be seen in Table~\ref{t_mp} and Fig.\ \ref{mp}, the pattern found
for the tangential velocity (\vtan) is symmetric with respect to the location of the
proposed outflow source, which is located close to the knot HH~223-K1.

The \vtan\ derived are quite similar for  both (blue- and redshifted)
outflow lobes. For the knots located on the  blueshifted CO outflow lobe
(\ie\ to the east of the millimetre sources) \vtan\ ranges
from $\sim$ 11 to 25 \kms\, with a mean velocity of 15.9 \kms. For the  knots
on the redshifted CO outflow lobe (to the west of the
millimetre sources) \vtan\ ranges from $\sim$ 4 to 32 \kms\, with a mean
velocity of 15.5 \kms.

\citet{Zha13} carry out a survey of H$_2$ outflows driven by low-mass
protostars in the L1688 core of $\rho$ Ophiuchi, and derive the proper
motions for 86 H$_2$ emission features. They find values in the range of
0.014--0.247~arcsec~yr$^{-1}$, which correspond to \vtan\ in the range of 
8--140~\kms, with a median velocity of 34.5~\kms. The proper motions we 
measured for the H$_2$ features of the HH~223 outflow lie within
this range, near its lower end. It should be noted that these authors also
find an appreciable drop in the number of features having low \vtan\ values,
which they mainly attribute to a bias effect. Thus, the \vtan\ we derived
for the H$_2$ features of the HH~223 outflow  are consistent with
typical values found in other H$_2$ outflows driven by low-mass protostars.

Regarding the direction of the knot proper motions, the  position
angles (PA) derived show that the knots located to the east of SMA~2 
move eastwards, in a direction opposite to that of the knots
located to the west of SMA~2, which  move westwards. In general, the
knot proper motions are projected following directions that are consistent,
within errors, with the direction of  the east-west CO outflow axis 
(PA of 110--115$\degr$, see \eg\ \citealp{Mor89}; \citealp{Lee02}), with
a few exceptions. The  knots closest ($\leq$ 10~arcsec east)  to SMA~2
(HH~223-K1 and -K2)  also move  eastwards  but the
PA of their proper motions are misaligned with respect to the large-scale
outflow
axis ($\Delta$ PA 30$\degr$-- 65$\degr$). However, at small scales, the ouflow
traced by the SiO emission has a PA close to that of the H$_2$ emission
of knots K1 and K2 (see \citealp{Gir09}; \citealp{Lop10}).
The proper motion of the knots H2-NW2c and H2-NW2d, $\sim$ 95
arcsec to  the west of the millimetre sources, are also misaligned  with
respect to the east-west CO outflow axis.  The CO velocity  maps of
\citet{Lee02} show the presence of a clump of blueshifted CO emission,
superposed on the redshifted CO outlfow lobe, around the location of these
H$_2$ knots. It might be indicative of a  complex interaction between molecular
clumps  at these positions that deviates the knot proper motions away from
the outflow axis. Such  deviations, attributed to complex interactions of
the outflow with a inhomogeneous environment, have been observed in other
jets (\eg\ HH~110, see \citealp{Kaj12}).

\subsection{Full spatial velocities}

\begin{figure}
\includegraphics[width=84mm,clip]{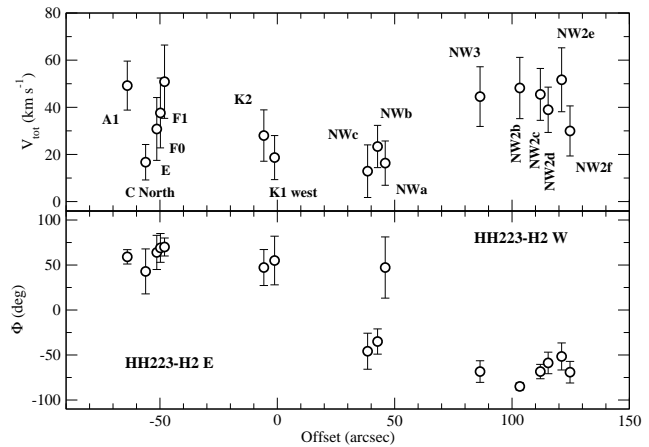}
\caption{Full spatial velocity (\vtot) and  angle ($\phi$) between the knot velocity
and the sky plane (with 
$\phi$ $\geq$ 0 towards the observer)
\label{v3d}} 
\end{figure}

The full spatial velocity (\vtot) and the angle ($\phi$) between the knot motion
and the sky (with  $\phi$ $\geq$ 0 towards the observer) was obtained for the
knots for which the two velocity components  
could be derived from the MOS and narrow-band imaging observations. 
Results are given in 
Table \ref{vtotfi} and plotted in Fig. \ref{v3d}.

The full spatial velocities derived for these knots range from
$\sim$ 15 to 55 \kms. No clear behaviour as a function  of the distance to SMA~2 is found. 
However, a trend can be seen, with  \vtot\ increasing  with distance from the exciting
source. Near SMA~2, the  knots  lying from $\sim$ --5 to +40 arcsec of SMA~2 have lower
\vtot (with a mean velocity of 17.8 \kms) than the knots far away of SMA~2,  to both sides of it.
Furthermore, \vtot\ is  slightly higher for the western (redshifted) knots lying at distances from +80 to
+120 arcsec, (with a mean velocity of 43 \kms), than for the eastern (blueshifted) knots lying at
distances from --10 to --70 arcsec to SMA~2 (with a mean velocity of 35.5 \kms).

In contrast, the inclination angle with respect to the plane
of the sky derived for the knot motions shows a  bipolar behaviour, as can
be appreciated in Fig.\ \ref{v3d}.  For the knots located to the east of
SMA~2, we derived inclination angles $\geq$ 50$\degr$, with a mean inclination angle 
of 58$\degr$. The motion of the knots lying to the west of SMA~2 is found
projected in the opposite direction with respect to the plane of the sky,
with a mean inclination angle of --60$\degr$. 

\begin{table}
\caption{Full spatial  velocities (\vtot) and inclination
angle ($\phi$)$^{1}$} 
\begin{tabular}{lrr}
\hline
Knot & \vtot & $\phi$ \\
HH~223- & (\kms)&   (degrees)\\
\hline
A1       &$ 49.2\pm 10.4$          &$ 59\pm\phantom{0}8$\\
C-North  &$ 16.7\pm\phantom{0}7.5$ &$ 43\pm25$\\
E        &$ 30.8\pm 13.3$          &$ 64\pm19$\\
F0       &$ 37.6\pm 15.0$          &$ 69\pm16$\\
F1       &$ 50.9\pm 15.6$          &$ 70\pm10$\\
K2       &$ 28.0\pm 10.9$          &$ 47\pm20$\\
K1-West  &$ 18.7\pm\phantom{0}9.3$ &$ 55\pm27$\\
H2-NWc   &$ 12.9\pm 11.2$          &$-46\pm20$\\
H2-NWb   &$ 23.4\pm\phantom{0}9.0$ &$-35\pm14$\\
H2-NWa   &$ 16.3\pm\phantom{0}9.4$ &$ 47\pm34$\\
H2-NW3   &$ 44.5\pm 12.7$          &$-68\pm12$\\
H2-NW2b  &$ 48.2\pm 12.0$          &$-85\pm\phantom{0} 4$\\
H2-NW2c  &$ 45.5\pm 11.0$          &$-68\pm\phantom{0} 8$\\
H2-NW2d  &$ 38.9\pm\phantom{0}9.6$ &$-59\pm12$\\
H2-NW2e  &$ 51.7\pm 13.6$          &$-52\pm15$\\
H2-NW2f  &$ 30.0\pm 10.6$          &$-69\pm12$\\ 
\hline 
\end{tabular}
\label{vtotfi}
\begin{list}{}
\item
$^{1}$ Angle between the knot velocity and the plane of the sky, positive 
towards the observer.
\end{list}
\end{table} 

\subsection{A scenario derived from the kinematics}

A scenario was previously outlined by \citet{Lop10}, which was based only on
the morphology of the near-infrared and CO outflows, with the
low-mass protobinary system VLA~2 lying  embedded in the millimetre source SMA~2. 
One of the protobinary members (most probably VLA 2A) is ejecting
supersonic gas at varying speed or
with different ejection directions, giving rise to the large-scale, east-west 
bipolar CO outflow.
In turn, the near-infrared HH~223 outflow is tracing slow shocks, excited by the
interaction of the CO outflow with the accelerated gas of the walls of the cavity
opened by it.  At some places,  the strength of the shocks increases, either
by interaction between parcels of  the outflow (internal working surfaces) or
with dense clumps of the cavity wall. These sites of stronger shocks are
traced by the \fe\ emission in the {\it J} and {\it H} spectral range,
provided that the extinction allows  the emission to emerge. The near-infrared
emission arising from the shocked gas has, in addition,  optical counterparts
at the regions with low visual extinction, giving rise to the Herbig-Haro
object 223, at the southeast edge of the outflow, and to the weak, diffuse 
HH~223-NW2 filament at its northwest edge, both structures being detected in
the H$\alpha$ line.

In this work, the full kinematics derived for the H$_2$ emission clearly 
confirms
that the location of the exciting outflow source has to be found around
the position of the millimetre source SMA~2, where the protobinary system is
embedded. A bipolar pattern has been found for the spatial distribution of
the  knots' proper-motion directions, as well as for the inclination angle
of the full velocity with respect to the plane of sky. In both cases, the
vectors  point in opposite direction, away from the region where the proposed
engine of the outflow is located. Furthermore, the radial velocity derived
along the near-infrared emission sampled with MOS changes  sign 
somewhere to the northwest of  HH~223-K1. Note that the 
location of the 
proposed engine of the outflow lies a few arcsec to the
northwest of HH~223-K1. Unfortunately  the high extinction prevented us to
detect near-infrared emission just coinciding with 
the position of the millimetre source SMA~2. Hence, from this work
it is not possible to establish 
which component of the protobinary system  is  closer  
to the location where
the velocity
 changes sign. 
 To address this subject, we would need to study the kinematics of the
emission in the neibourghood of HH~223-K1 using tracers of the emission
from dense gas that would allow us to penetrate  closer to the
millimetre source.

Finally, we found some departures from the general trend followed by the
kinematics of the near-infrared emission at several positions. We
interpreted them as caused by a more complex interaction with an
inhomogeneous, clumpy environment or between different
parcels of the outflow.

\section{SUMMARY AND CONCLUSIONS}

Taking advantadge of the Multi-Object-Spectroscopy (MOS)
observing mode of LIRIS, we obtained {\it J}, {\it H} and {\it K}-band spectroscopy of the 
complex morphology of the L723 outflow.
We obtained the full kinematics of the near-infrared (H$_2$) outflow HH~223, located in the
L723 dark cloud. The proper motions were derived from multi-epoch H$_2$ 
images of the L723 field, obtained through a narrow-band filter centred on  the 2.122~$\mu$m  line. 
The radial velocities were derived from the 2.122~$\mu$m H$_2$ lines of the MOS spectra.
Hence, both radial and tangential velocities  correspond to the same 
emitting outflow gas.
The kinematics derived from the data presented in this work lead us to the conclusions
summarized below.

\begin{itemize}

\item
At large scales, the radial velocity shows a bipolar pattern along the
outflow, with negative (blueshifted) values towards
the southeast, and positive (redshifted) values  towards the northwest
from the position of SMA~2. The spatial 
distributions of the radial velocity of  the H$_2$ and CO emissions match each other: 
the H$_2$ knots having negative velocity lie
on the blueshifted CO lobe, while the H$_2$ knots with positive velocity lie
on the redshifted CO lobe. This strongly supports 
that the H$_2$ and CO
outflows share their exciting source.

\item

The proper motions derived lie within the range of values found in other H$_2$ outflows
driven by low-mass protostars. The proper motions of  the blueshifted and of the
redshifted H$_2$ knots have similar values. As a general trend, the knots move
following the direction of the CO outflow axis. The proper motions follow a bipolar pattern,
 with the knots to the east of SMA~2 moving in a direction opposite to the motions of the
knots to the west of SMA~2.

\item

For the brightest H$_2$ knots, we derived the full spatial velocity 
and the inclination angle of their 
motion with  respect to the plane of the sky. 
A bipolar pattern, centred around the location of SMA~2 
(where the exciting outflow source is embbeded), is found: the eastern knots 
move towards the
observer ($\phi$~$\simeq$~+60$\degr$), while the   western knots 
move far away to the observer ($\phi$~$\simeq$~--60$\degr$).

\item
In addition, we built the PV maps of the 1.748~$\mu$m line, the brightest H$_2$  line 
detected in the observed {\it H-} spectral range, and derived the  radial velocity 
from this
line. 
Both, radial velocities and spatial brightness distribution are 
consistent with those
derived from the 2.122~$\mu$m line.  

\item

Emission from \fe\ lines, tracing the emission of the ionized gas, was
only detected in the  {\it J-} and {\it H-} band spectra of the slitlets
positioned on HH~223-A to HH~223-F. Since these knots are far away from the
location of the powering outflow source, the \fe\ emission has to be
originated in shocks. 

\end{itemize}

In summary, the 3D kinematics of the  HH~223 H$_2$ outflow derived in this
work confirms that all the H$_2$ nebular structures form part of a large-scale,
S-shaped near-infrared outflow, which is powered by the YSO located 
within the millimetre source SMA~2, the closest one to the H$_2$ knot
HH~223-K1. The kinematics also confirms that the H$_2$ and CO outflows
are physically related.

\section*{Acknowledgments}
The paper is based on observations made with the  4.2~m William Herschel
Telescope operated on the island of La Palma by the ING at the Observatorio del
Roque  de los Muchachos of the Instituto de Astrof\' \i sica de Canarias.\\ 
J. A-P. is partially supported by  AYA2011-25527. B. G.-L. acknowledges
finantial support from the Spanish MINECO grants AYA2012-39408-C0-02 and
AYA2013-41656-P. 
R. E. and R. L. are partially supported by the Spanish MICINN grant
AYA2011-30228-C03-03. 
R. L. akcnowledges the hospitality of the Instituto de Astrof\'isica de
Canarias, where part of this work was done.\\
We thanks the referee, Prof. A. Raga, for the detailed revision of the manuscript.

\label{lastpage}
\end{document}